# Globally Coupled Maps: Almost Analytical Treatment of Phase Transitions

Wolfram Just
Theoretische Festkörperphysik
Technische Hochschule Darmstadt
Hochschulstraße 8
D–64289 Darmstadt
Germany

May 3, 1994

**Abstract**

Bifurcations in a system of coupled maps are investigated. Using symbolic dynamics it is proven that for coupled shift maps the well known space–time–mixing attractor becomes unstable at a critical coupling strength in favour of a synchronized state. For coupled non–hyperbolic maps analytical and numerical evidence is given that arbitrary small coupling changes the dynamical behaviour. The anomalous dependence of fluctuations on the system size is attributed to these bifurcations.

# 1 Introduction

The influence of chaotic dynamics in spatially extended systems is a field of intensive research. The competition between local chaos and diffusive coupling seems to be at the heart of pattern formation out of a random state. A detailed knowledge of this mechanism seems to be necessary to understand the development of structures out of regular and irregular states (cf. [1] and references cited therein). These concepts link different fields like e.g. hydrodynamic, optical and magnetic instabilities, chemical reactions or biological systems. It is usually tremendous complicated to gain information from the basic equations of motion even if numerical simulations are concerned. For that reason on is forced to investigate simple model systems, a strategy which has proven to be fruitful in several fields of low dimensional chaotic systems [2, 3]. One class of suitable model systems are constituted by coupled maps which are known to incorporate many features of real time evolution from a phenomenological point of view (e.g. [4]). They are accessible not only by numerical methods with moderate effort but also allow for a partially rigorous treatment. Especially the weak coupling limit of hyperbolic coupled maps has been considered recently [5, 6, 7]. It has been shown that even in the limit of infinite system size the dynamics is dominated by exponentially decaying correlation functions, a property which is usually termed space–time–mixing. The motion is similar to the dynamics of the uncoupled system. Indeed it has been shown that the map lattice can be decoupled by a continuous transformation. Furthermore it has been suggested that pattern formation can be related to equilibrium phase transitions in higher dimensional spin systems using a thermodynamical formulation via symbolic dynamics. However no definite results are available at the moment. Furthermore (partially) solvable model systems are lacking which are suitable to study these problems.

It is the aim of this paper to investigate a simple coupled map lattice beyond the weak coupling regime and to get as far as possible rigorous results. To be definite a lattice of maps on the circle $S^1 = [0, 2\pi]$ is considered. The circle and not the interval is chosen as phase space because hyperbolicity, the prerequisite to obtain rigorous results, is guaranteed for local expanding maps [8]. The equations of motion read

$$\varphi_{n+1}^{(\nu)} = \left(T(\underline{\varphi}_n)\right)^{(\nu)} = f(\varphi_n^{(\nu)}) + \frac{\epsilon}{L}\sum_{\mu} g\left(f(\varphi_n^{(\mu)}) - f(\varphi_n^{(\nu)})\right) \ |\operatorname{mod} 2\pi \quad . \tag{1}$$

$L$ denotes the lattice size, $\epsilon$ the strength of the coupling, and $\underline{\varphi} = (\varphi^{(\nu)}) \in S^L$ the phase space coordinates. The map $f$ which determines the single site dynamics is supposed to be given by

$$f(\varphi) = 2\varphi + a\sin(\varphi) \ |\operatorname{mod} 2\pi \quad . \tag{2}$$

For different values of the parameter $a \in [0, 2]$ it interpolates between the simple case of the shift map ($a = 0$) and a strongly non–hyperbolic map which has a critical point ($a = 2$). In the latter case it resembles strongly the structure of the Smale complete logistic equation. Continuity of the map $T$ requires periodicity for the coupling function $g$. For explicit calculations the simplest choice

$$g(x) = \sin(x) \tag{3}$$

will be considered. A global coupling between the different lattice sites has been assumed in eq.(1). This type of coupling considerably simplifies a theoretical approach. It has also been suggested that such a mean field approach yields the correct description of physically more reasonable models with short range coupling above some critical dimension in accordance to equilibrium statistical mechanics [4, 9]. Furthermore globally coupled models are interesting in their own right. They may be considered on one hand as limiting cases of models with long range interactions [10]. On the other hand they have been considered in the context of biology, neural networks and Hamiltonian dynamics of coupled oscillators [11, 12, 13, 14]. Special emphasis has been laid on the mechanism of synchronization between the different elements. Analogous phenomena can be studied in the model (1). In contrast to the mentioned approaches the randomness is not modeled by a stochastic force but is an inherent property of the dynamical system.

The plan of the paper reads as follows. Section 2 reviews the well known construction of a symbolic dynamics for the coupled map lattice (1) on an elementary level. With its help a space–time–mixing stationary state can be established for sufficiently small coupling strength $\epsilon < \epsilon_c^e$. In the subsequent sections the two limiting cases of hyperbolic coupled maps ($a = 0$) and strongly non–hyperbolic coupled maps ($a \lesssim 2$) are discussed separately. In section 3 it is shown rigorously that the breakdown of the space–time–mixing regime is accompanied by a global synchronization among the lattice sites for moderate coupling strength $\epsilon_c^e < \epsilon < \epsilon_c^m$. Furthermore space–time chaotic transients of tremendous length are observed which are attributed to a hyperbolic repeller. The case of non–hyperbolic coupled maps is much more difficult to analyse. In section 4 analytical and numerical evidence is given that even for infinitesimal coupling strength complicated bifurcations arise. They cause an anomalous dependence of the fluctuations on the system size, an effect which has been commonly observed in globally coupled non–hyperbolic maps and has been termed "violation of the law of large numbers" [15, 16, 17, 18]. Finally comments on prospective work are given.

# 2 Markov partitions for coupled map lattices

It is the objection of this paragraph to review the well known results on the symbolic dynamics of system (1) on an elementary level [5]. Additionally I will set up some notation which will be useful in the subsequent discussion.

To construct the symbolic dynamics only some global features of the single site map $f$ and the coupling $g$ is needed but not the explicit expressions (2) and (3). Especially one demands that the single site map is monotonous. It ensures that $f$ admits a Markov partition[1] [8]. For simplicity in notation let us assume a binary partition which is written as $I_0 = [0, \pi]$, $I_1 = [\pi, 2\pi]$ without loss of generality. $f$ maps these intervals to the whole phase space in a monotonous way, $f(I_i) = [0, 2\pi]$. With this partition a symbol sequence $(\sigma_0, \sigma_1, \ldots)$, $\sigma_i \in \{0, 1\}$ is assigned to every phase space point $\varphi$. It denotes the element $I_{\sigma_k}$ of the partition that contains the image point $f^k(\varphi)$, $k \geq 0$ [2]. For further simplification in the notation let us demand that the coupling obeys $g(0) = 0$ and $\max\{g'(\varphi)\} = g'(0) = 1$. Obviously the choice (2) and (3) shares all these properties.

The Markov partition of the uncoupled map lattice, $\epsilon = 0$ can be obtained trivially as a direct product $U_{\underline{\sigma}} = I_{\sigma^{(0)}} \otimes \cdots \otimes I_{\sigma^{(L-1)}}$, $\underline{\sigma} = (\sigma^{(0)}, \ldots, \sigma^{(L-1)})$. The boundaries of these sets are given by the "hyperplanes" $\varphi^{(\nu)} = 0$ and $\varphi^{(\nu)} = \pi$ respectively. The elements of this partition are mapped by $T$ to the whole phase space in a bijective way. If the single site map is expanding then the partition is also generating. Any sequence $\underline{\sigma}_0, \underline{\sigma}_1, \ldots$ specifies a phase space point via $\bigcap_{k\geq 0} T^{-k}(U_{\underline{\sigma}_k})$. The dynamics is equivalent to a shift operation on the two dimensional spin lattice $\underline{\sigma}_0, \underline{\sigma}_1, \ldots$.

The Markov partition obtained for the uncoupled case can be carried over to the case of finite coupling. The discussion is somewhat simplified if one considers the extended map $T : \mathbb{R}^L \to \mathbb{R}^L$ which develops from the original evolution equation (1) by suppressing the modulo operation

$$x_{n+1}^{(\nu)} = \left(\tilde{T}(\underline{x}_n)\right)^{(\nu)} = \tilde{f}(x_n^{(\nu)}) + \frac{\epsilon}{L} \sum_{\mu} g\left(\tilde{f}(x_n^{(\mu)}) - \tilde{f}(x_n^{(\nu)})\right) \quad . \tag{4}$$

Although this system has orbits which tend to infinity the original dynamics is recovered via $\varphi_n^{(\nu)} = x_n^{(\nu)}|\text{mod}2\pi$. Furthermore any object in the original phase space $S^L$ (e.g. the partition $U_{\underline{\sigma}}$) can be identified with the corresponding object in the extended space $\mathbb{R}^L$ by $2\pi$ periodic continuation. The action of eq.(1) respectively (4) is locally invertible for moderate coupling. Using the linearization

$$\left(D\tilde{T}(\underline{x})\right)_{\nu\mu} = \tilde{f}'(x^{(\mu)}) \left[\delta_{\nu\mu} - \frac{\epsilon}{L} \sum_{\rho} g'(\tilde{f}(x^{(\rho)}) - \tilde{f}(x^{(\nu)}))\delta_{\nu\mu} + \frac{\epsilon}{L} g'(\tilde{f}(x^{(\mu)}) - \tilde{f}(x^{(\nu)}))\right] \tag{5}$$

one obtains for the determinant (cf. appendix A)

$$\det(D\tilde{T}(\underline{x})) \;=\; \prod_{\nu=0}^{L-1} \tilde{f}'(x^{(\nu)}) \left[\prod_{\nu} \left(1 - \frac{\epsilon}{L} \sum_{\mu} \gamma_{\nu\mu}\right) + \frac{\epsilon}{L} \sum_{\nu} \prod_{\rho(\neq\nu)} \left(1 - \frac{\epsilon}{L} \sum_{\mu} \gamma_{\rho\mu}\right)\right.$$

[1] If additionally $f$ is locally expanding, $|f'| > 1$ then the partition is generating.

[2] More precisely: $\varphi \in \bigcap_{k\geq 0} f^{-k}(I_{\sigma_k})$ yields a homomorphism between the dynamics and the shift of symbol sequences.

$$+\frac{1}{2}\left(\frac{\epsilon}{L}\right)^2 \sum_{\nu\neq\rho}\left(1-\gamma_{\nu\rho}^2\right)\prod_{\mu(\neq\nu,\rho)}\left(1-\frac{\epsilon}{L}\sum_{\sigma}\gamma_{\mu\sigma}\right)\Bigg] \qquad (6)$$

where the abbreviation $\gamma_{\nu\mu} = g'(\tilde{f}(x^{(\mu)}) - \tilde{f}(x^{(\nu)}))$ has been used. As $\gamma_{\nu\mu} \leq 1$ and $f' > 0$ by definition it follows that

$$\det(D\tilde{T}(\underline{x})) \geq \prod_{\nu=0}^{L-1} \tilde{f}'(x^{(\nu)}) \left[(1-\epsilon)^L + \epsilon(1-\epsilon)^{L-1} + \frac{1}{2}\epsilon^2 \frac{L-1}{L}(1-\epsilon)^{L-2}\right] \quad . \qquad (7)$$

Hence the determinant does not vanish for $0 \leq \epsilon < \epsilon_c^m = 1$ which proves local invertibility. In contrast to the map (1) the extended system (4) is in addition globally invertible because its phase space is simply connected. This property yields the main difference between both formulations.

We recall the fact that the boundaries of the Markov partition are given by the hyperplanes $x^{(\nu_0)} = 0$ respectively $x^{(\nu_0)} = \pi$ in the uncoupled case $\epsilon = 0$. They are the preimages of the hyperplanes $x^{(\nu_0)} = 0$ and $x^{(\nu_0)} = 2\pi$ which separate the extended phase space according to $2\pi$ periodicity. It is this property which guarantees that the elements of the partition $U_{\underline{\sigma}}$ are mapped to the whole phase space $S^L$. This observation is used to construct the partition for finite coupling $0 < \epsilon \leq \epsilon_c^m$. To this end let us consider for every fixed lattice site $\nu_0 \in \{0, \ldots, L-1\}$ the preimages of the hyperplanes $x^{(\nu_0)} = 0$ respectively $x^{(\nu_0)} = 2\pi$. They are determined by the equations

$$\begin{aligned} 0 &= H_{x^{(\mu)},\mu\neq\nu_0}(x^{(\nu_0)}) := f(x^{(\nu_0)}) + \frac{\epsilon}{L}\sum_{\mu} g\left(f(x^{(\mu)}) - f(x^{(\nu_0)})\right) \\ 2\pi &= H_{x^{(\mu)},\mu\neq\nu_0}(x^{(\nu_0)}) \quad . \end{aligned} \qquad (8)$$

For every choice $x^{(\mu)}, \mu \neq \nu_0$ they yield a unique solution $x^{(\nu_0)}$ because $H'_{x^{(\mu)},\mu\neq\nu_0}(x^{(\nu_0)}) > 0$ holds for $0 \leq \epsilon < \epsilon_c^m$. Therefore they define codimensions 1 manifolds in the phase space and slice the space in "hypercubes". These sets are mapped by $\tilde{T}$ to a full cube $[0, 2\pi]^L$ in a bijective way. Fig.1 shows this construction in the simple case $L = 2$. < Fig.1
The cubes can be labeled uniquely by symbol sequences $\underline{\sigma}$ because they develop continuously from the partition of the uncoupled case. This explicit construction yields a partition $U_{\underline{\sigma}}$ of the phase space so that $T : U_{\underline{\sigma}} \to S^L$ is invertible. Hence $T^{-1}(U_{\underline{\sigma}_1}) \cap U_{\underline{\sigma}_0}$ is not empty and the set $\cap_{k\geq 0} T^{-k}(U_{\underline{\sigma}_k})$ contains at least one point for any sequence $\underline{\sigma}_0, \underline{\sigma}_1, \ldots$. Explicit expressions for the boundaries of the partition are given by eqs.(8).

To make the partition a generating one, that means that $\cap_{k\geq 0} T^{-k}(U_{\underline{\sigma}_k})$ contains exactly one point, one has to impose some expansion property. In such a case the diameters of the successive preimages $T^{-n}(U_{\underline{\sigma}_n}) \cap T^{-(n-1)}(U_{\underline{\sigma}_{n-1}}) \cap \ldots \cap U_{\underline{\sigma}_0}$ tend to zero. This property is obviously fulfilled if the map $\tilde{T}$ expands any points[3]. It is

[3]more precisely: the image of any $\delta$ neighbourhood of a point $\underline{x}$ contains a $\delta'$ neighbourhood of $\tilde{T}(\underline{x})$ with $\delta' > \delta$.

shown easily (cf. appendix B) that this global expansion property coincides with local expansiveness. It states that infinitesimal neighbouring points separate in the course of the dynamics

$$\forall \underline{y} \quad \|D\tilde{T}(\underline{x})\underline{y}\| \geq c\|\underline{y}\|, \quad c > 1 \quad . \tag{9}$$

Expansiveness may pose additional constraints on the coupling strength $\epsilon$. In the uncoupled case this property is obviously fulfilled for expanding maps, $|f'| > 1$. Hence it persists up to some critical value $\epsilon_c^e$. In the range $0 \leq \epsilon < \epsilon_c^e$ the dynamics is equivalent to a shift in a two dimensional spin lattice $(\underline{\sigma}_0, \underline{\sigma}_1, \ldots)$. This property proves that the coupled system can be decoupled by a continuous coordinate transformation[4]. The dynamics in this state is mixing with respect to time as well as space translations [5]. It is however difficult to compute the explicit value $\epsilon_c^e$ from eq.(9) without specifying the single site map and the coupling. So we postpone the evaluation to the next sections.

Let me close this paragraph by stressing briefly the opposite case of large coupling. Because of the infinite range of interaction any lattice sites which take the same phase space value, $\varphi_n^{(\nu)} = \varphi_n^{(\mu)}$, will maintain this property forever. Such a region of constant phase may be called a domain. The most trivial domain is the globally synchronized solution $\varphi_n^{(\nu)} = \phi_n$. The dynamics of such a solution is governed by the single site map $\phi_{n+1} = f(\phi_n)$. Its stability is determined by the linearized dynamics which in view of eq.(1) reads

$$\delta\varphi_{n+1}^{(\nu)} = f'(\phi_n) \left[ (1 - \epsilon g'(0))\delta\varphi_n^{(\nu)} + \frac{\epsilon}{L} g'(0) \sum_\mu \delta\varphi_n^{(\mu)} \right] \quad . \tag{10}$$

The solution can be written as

$$\delta\varphi_n^{(\nu)} = \exp\left( i\frac{2\pi}{L} k\nu \right) \delta a_n(k), \quad 0 \leq k \leq L-1 \tag{11}$$

where the amplitudes obey

$$\delta a_{n+1}(k) = \lambda_k f'(\phi_n)\delta a_n(k), \quad \lambda_{k=0} = 1, \quad \lambda_{k\neq 0} = 1 - \epsilon g'(0) \quad . \tag{12}$$

Stability requires that the coefficient on the right hand side is smaller than unity on the average for $k \neq 0$. This condition determines a critical coupling strength $\epsilon_c^s$ via

$$\exp\left(\langle \ln|f'(\varphi)| \rangle\right) (1 - \epsilon_c^s g'(0)) = 1 \tag{13}$$

where $\langle \cdots \rangle$ denotes the long time average. Synchronization sets in for $\epsilon > \epsilon_c^s$. Obviously $\epsilon_c^e \leq \epsilon_c^s$ holds. The transition from the space–time–mixing regime to the synchronized state will be at the center of interest of the next section.

[4] In generally this transformation seems to be of no practical use because of its complicated structure.

# 3 Hyperbolic coupled maps

The most simple case of coupled shift maps, $a = 0$ has been discussed in the literature from several point of views (e.g. for mathematical considerations [19], for the case $L = 2$ [20]). It allows for a complete analytical treatment for moderate coupling $0 < \epsilon < \epsilon_c^m$ which incorporates the transition to the synchronized state. The discussion of this transition is at the heart of this section.

First of all the condition of expansivity eq.(9) can be evaluated easily. Using eqs.(2), (3) and (5) for $a = 0$ one obtains for arbitrary vectors $\underline{y}$ in the tangent space

$$\begin{aligned}
\langle \underline{y} | D\tilde{T}(\underline{x}) | \underline{y} \rangle &= 2 \sum_\nu \left( y^{(\nu)} \right)^2 - \frac{2\epsilon}{L} \sum_{\nu,\rho} \cos(2x^{(\rho)} - 2x^{(\nu)}) \left( y^{(\nu)} \right)^2 \\
&\quad + \frac{2\epsilon}{L} \left| \sum_\mu \exp(i2x^{(\mu)}) y^{(\mu)} \right|^2 \\
&\geq (2 - 2\epsilon) \langle \underline{y} | \underline{y} \rangle \quad . \qquad (14)
\end{aligned}$$

Owing to the fact that the matrix (5) is symmetric for $a = 0$ the condition (9) follows from eq.(14) for $0 \leq \epsilon < \epsilon_c^e = 1/2$. It implies that the dynamics is equivalent to a shift operation in a two dimensional spin lattice $\underline{\sigma}_0, \underline{\sigma}_1, \ldots$ and is mixing with respect to the time evolution as well as space translations[5].

On the other hand the stability condition for the synchronized state (13) immediately yields a local stable uniform solution for $\epsilon_c^s = 1/2 < \epsilon$. Hence the breakdown of space–time–mixing is related to a global synchronization in the system. The natural question arises what causes the transition from the space–time–mixing regime to the synchronized state and whether there exist other stable solutions.

Before we enter the analysis of the general case let us consider first the simple situation of two coupled shift maps. It contains the essence of the subsequent analysis on an elementary level. If we use the extended description (4) and introduce symmetric coordinates $x^{(\pm)} = x^{(1)} \pm x^{(0)}$ the system can be decoupled

$$\begin{aligned}
x_{n+1}^{(+)} &= 2x_n^{(+)} \\
x_{n+1}^{(-)} &= 2x_n^{(-)} - \epsilon \sin\left(2x_n^{(-)}\right) \quad . \qquad (15)
\end{aligned}$$

As the variables $x^{(\nu)}$ are of interest only modulo $2\pi$ the system (15) can be considered modulo $2\pi$ respectively $4\pi$. The first equation describes the chaotic motion parallel to the diagonal of the phase space. The dynamics perpendicular to this direction is governed by the second one. The corresponding map is depicted in Fig.2. For $0 \leq \epsilon < \epsilon_c^s$ the map possesses a slope larger than unity and yields a chaotic attracting set. Approaching $\epsilon_c^s$ from below intermittency near the fixed point $x^{(-)} = 0$ sets in. < Fig.2

[5]On a rigorous mathematical level the latter statement requires additional investigations [5].

Above the critical value the fixed point becomes stable via a pitchfork bifurcation. Furthermore there remains an invariant repelling Cantor set in the region bounded by the two unstable fixed points. The transition is usually termed a boundary crisis [21]. The fixed point is globally stable but chaotic transients occur.

The dynamics of the general system (4) can be treated in a quite similar fashion for $\epsilon_c^s < \epsilon < \epsilon_c^m$. To this end let us consider a neighbourhood of the diagonal[6]

$$\tilde{A}_\delta = \{\underline{x} \in \mathbb{R}^L \,|\, |x^{(\nu)} - x^{(\mu)}| \leq \delta\} \quad . \tag{16}$$

In a first step the diameter $\delta$ is chosen in such a way that the system (4) contracts the neighbourhood towards the synchronized state. For $\underline{x}_n \in \tilde{A}_{\delta_*}$ we obtain

$$\begin{aligned} |x_{n+1}^{(\mu)} - x_{n+1}^{(\nu)}| &= \left| 2 - \frac{2\epsilon}{L} \sum_\rho \cos(2x_n^{(\rho)} - x_n^{(\nu)} - x_n^{(\mu)}) \frac{\sin(x_n^{(\mu)} - x_n^{(\nu)})}{x_n^{(\mu)} - x_n^{(\nu)}} \right| |x_n^{(\mu)} - x_n^{(\nu)}| \\ &= \left| 2 - \frac{2\epsilon}{L} \left[ 2\cos(x_n^{(\mu)} - x_n^{(\nu)}) + \sum_{\rho(\neq \nu,\mu)} \cos(2x_n^{(\rho)} - x_n^{(\mu)} - x_n^{(\nu)}) \right] \right. \\ &\quad \left. \cdot \frac{\sin(x_n^{(\mu)} - x_n^{(\nu)})}{x_n^{(\mu)} - x_n^{(\nu)}} \right| |x_n^{(\mu)} - x_n^{(\nu)}| \\ &\leq \left| 2 - 2\epsilon \left[ \frac{2}{L} \cos(\delta_*) \frac{\sin(\delta_*)}{\delta_*} + \frac{L-2}{L} \cos(2\delta_*) \right] \right| |x_n^{(\mu)} - x_n^{(\nu)}| \quad . \end{aligned} \tag{17}$$

Choosing $\delta_*$ sufficiently small the prefactor becomes smaller than unity and the map contracts the neighbourhood $\tilde{A}_{\delta_*}$ for $\epsilon_c^e < \epsilon < \epsilon_c^m$

$$2 - 2\epsilon \left[ \frac{2}{L} \cos(\delta_*) \frac{\sin(\delta_*)}{\delta_*} + \frac{L-2}{L} \cos(2\delta_*) \right] < 1 \quad . \tag{18}$$

It is obvious that the same property holds for the corresponding neighbourhood $A_{\delta_*}$ of the diagonal in the phase space $S^L$. In terms of Fig.2 this set is contained in the interval whose final points are given by the two unstable fixed points.

In a second step let us choose $\delta_*$ in such a way that the map $T$ locally expands distances on the complement of $A_{\delta_*}$. For $\underline{\varphi} \in A_{\delta_*}^C$ and arbitrary vector $\underline{y}$ in the tangent space we obtain

$$\begin{aligned} \langle \underline{y} | D\tilde{T}(\underline{\varphi}) | \underline{y} \rangle &= 2 \sum_\nu \left( y^{(\nu)} \right)^2 - \frac{\epsilon}{L} \sum_{\nu,\mu} \cos(\varphi^{(\mu)} - \varphi^{(\nu)}) \left( y^{(\nu)} - y^{(\mu)} \right)^2 \\ &\geq 2 \sum_\nu \left( y^{(\nu)} \right)^2 - 2\frac{\epsilon}{L} \cos(\delta_*) \left[ L \sum_\nu \left( y^{(\nu)} \right)^2 - L \left( \sum_\nu y^{(\nu)} \right)^2 \right] \\ &\geq [2 - 2\epsilon \cos(2\delta_*)] \langle \underline{y} | \underline{y} \rangle \quad . \end{aligned} \tag{19}$$

[6] Only for technical reasons the subsequent discussion is performed using the extended formulation. Let me stress again that $A_\delta$ denotes the corresponding set in the phase space $S^L$.

Hence the map is expansive on $A_{\delta_*}^C$ if the condition

$$2 - 2\epsilon\cos(2\delta_*) > 1 \tag{20}$$

is valid. This set corresponds in Fig.2 to the region where the map has slope larger than 1.

Above the critical coupling strength $1/2 = \epsilon_c^e = \epsilon_c^s < \epsilon$ both relations (18) and (20) can be satisfied by some (non–unique) value $\delta_*$. As a consequence the map (1) contracts on $A_{\delta_*}$ towards the diagonal and is expanding on the complementary set. It remains to investigate the invariant set which is contained in this part of the phase space. To this end we use again the Markov partition $U_{\underline{\sigma}}$ which is available for $0 \leq \epsilon \leq \epsilon_c^m = 1$. The sets $V_{\underline{\sigma}} = U_{\underline{\sigma}} \cap A_{\delta_*}^C$[7] yield a partition of the set under investigation. Because $T : U_{\underline{\sigma}} \to S^L$ is invertible and $T(A_{\delta_*}) \subset A_{\delta_*}$ holds it follows that $T(V_{\underline{\sigma}}) \supset A_{\delta_*}^C$ and $T$ is invertible on its image (more precisely: $T : V_{\underline{\sigma}} \to T(V_{\underline{\sigma}})$ is invertible). Therefore $T^{-n}(V_{\underline{\sigma}_n}) \cap \ldots V_{\underline{\sigma}_0}$ is not empty for any symbol lattice $\underline{\sigma}_0, \underline{\sigma}_1, \ldots, \underline{\sigma}_n$. Owing to the expansivity of the map $T$ on $A_{\delta_*}^C$ the set $\cap_{k\geq 0} T^{-k}(V_{\underline{\sigma}_k})$ contains exactly one point of the invariant set. The dynamics on this set is again equivalent to a shift in the two dimensional spin lattice $\underline{\sigma}_0, \underline{\sigma}_1, \ldots$. Finally let me mention that the invariant set contains no open neighbourhood because of the expansivity of the map (cf. appendix B). Therefore the dynamics beyond the critical value $\epsilon_c^s$ is determined by a globally stable synchronized state and a repelling space–time–mixing invariant set.

Let us now study how the invariant sets influence the time evolution of the coupled map lattice. Below the critical coupling strength $\epsilon_c^s$ the space–time–mixing state yields a quite trivial dynamics. The numerical simulations show a very noisy behaviour. Therefore let me focus on the regime $\epsilon_c^s < \epsilon$. To visualize the time evolution of numerical simulations some global quantity is useful. Here the absolute value of the "mean field" $r = |\sum_\nu \exp(if(x^{(\nu)}))/L|$ (cf. section 4) has proven to be a suitable quantity. Its value fluctuates in the space time mixing regime and saturates to 1 in the synchronized state. A few typical time series are shown in Fig.3 for coupling strength $\epsilon = 0.65, 0.8$ and systems size $L = 15, 21$. They have been obtained from a randomly chosen initial condition. The time series split into a random transient whose duration seems to increase with the system size and a sharp relaxation towards the synchronized solution. The sharpness of the transition enables us to introduce a well defined relaxation time < Fig.3

$$N_A(\underline{\varphi}) = \min\{n \in \mathbb{N}, | T^n(\underline{\varphi}) \in A\} \tag{21}$$

depending on the initial condition $\underline{\varphi}$ and some neighbourhood $A$ of the synchronized state (e.g. the set (16)). Because of the fast relaxation this quantity is practically independent of the choice of $A$ as far as this set is contained in the attracting region (16). The relaxation time fluctuates strongly. Its distribution is given by

$$P_n(A) = \langle \delta(n, N_A(\underline{\varphi})) \rangle \tag{22}$$

where $\langle \ldots \rangle$ denotes an average over the distribution of initial points. The quantity (22) has been simulated numerically for various system lengths and coupling strength by taking an average over 10000 randomly chosen initial conditions. The relaxation time, that means the set $A$, has been defined by the condition that the mean field deviates from the final value 1 by less than $10^{-3}$. Fig.4 shows some representative results. The distribution is in all cases Poisson like with an exponential tail. The least square fit between the maximum of the distribution and the first box that contains only one member of the ensemble is indicated also. The slope of this fit yields the tail of the distribution $P_n(A) \simeq C_1 \exp(-\lambda_{\epsilon,L} n)$ and coincides with the inverse of the mean relaxation time. The $\epsilon$ and $L$ dependence of this quantity has been investigated from the numerical data. Fig.5 contains the dependence on the system size for various coupling strengths. The numerical capabilities restrict the analysis to small systems or coupling strength far above the critical value. However in the whole range a exponential dependence on the system size is clearly indicated, $\lambda_{\epsilon,L} \simeq \exp(-L\alpha_\epsilon)$. From this observation it is evident that large systems yield huge transient times. The slopes in Fig.5 allow for an analysis of the $\epsilon$ dependence also. The evaluation suggests a logarithmic dependence (cf. Fig.6) , $\alpha_\epsilon \simeq -\ln C_2 - a \ln(\epsilon - \epsilon_c^s)$, with a slope $a \sim 0.4$. Summarizing the numerical findings we have observed a Poisson like distribution of relaxation times where the inverse of the mean relaxation time obeys $\lambda_{\epsilon,L} \simeq C_1[C_2(\epsilon - \epsilon_c^s)^a]^L$. Near the transition point $\epsilon_c^s$ and for large system size a huge space–time–mixing transient is observed.

[7] The sets $V_{\underline{\sigma}}$ are not empty because $U_{\underline{\sigma}}$ contains points of the form $\varphi^{(\mu)} = 0$, $\mu \neq \nu_0$, $\varphi^{(\nu_0)} = \pi$ by construction.

< Fig.4

< Fig.5

< Fig.6

The numerical observations can be explained in terms of the invariant sets discovered above. First of all the Poisson like distribution of relaxation times is closely related to the decay rate of the space–time–mixing repeller. In order o apply the conventional terminology of decay rates [22] it is however necessary that the invariant set has vanishing Lebesgue measure. Even in the simple case considered here this property seems to be difficult to proof which renders calculations from first principles impossible. If we however assume that the repeller has measure zero then the shape of the distribution is an immediate consequence of a finite decay rate which coincides with the rate $\lambda_{\epsilon,L}$ (cf. appendix C). The decay rate can be estimated by simple arguments without referring to tedious calculations. Owing to the hyperbolic structure and the Markov partition of the space–time–mixing state the motion on this set can be viewed as a stochastic process in phase space. The probability that a point falls on one step of iteration within the attracting region (16) of the synchronized solution is given by its volume relative to the whole phase space measure. It is simply estimated as $p \simeq 2\pi(\delta_*)^{L-1}/(2\pi)^L \sim (\delta_*/(2\pi))^L$. The probability that the dynamics settles on the synchronized state after $n$ iteration steps is simply given by the Binomial dis-tribution $P_n \simeq (1-p)^{n-1}p \sim \exp(-np)$ which reduces to a Poisson distribution for small probability $p$. Its decay rate $p$ yields the exponential dependence on the system size. The coupling strength enters via the diameter $\delta_*$ of the attracting region. From eq.(18) or (20) it can be immediately estimated as $\delta_* \sim (\epsilon - \epsilon_c^s)^{1/2}$ in reasonable agreement with the numerical data.

The transition from the space–time–mixing to the synchronized state at $\epsilon_c^s$ is accompanied by long transients which are caused by a space–time–mixing repeller with a small escape rate. The nature of this repeller makes it difficult to observe the transition in numerical simulations if the coupling is varied adiabatically. Furthermore for system size $L \gtrsim 30$ the transients may become longer than any reasonable observation time and no synchronization will be observed at all. In the thermodynamic limit $L \to \infty$ the space–time–mixing state becomes in a certain sense stable. The transition from the synchronized state to the space–time–mixing regime which is observed if one decreases the coupling strength is always a sharp transition because it is brought about by the local properties near the diagonal. Hence the synchronization is accompanied by a strong hysteresis effect.

# 4 Non–hyperbolic coupled maps

Let us now focus on the opposite case $a \lesssim 2$. At the beginning I should mention that to my best knowledge there are no general results available whether the single site map (2) is chaotic in this case. Nevertheless slight modifications of the approaches used e.g. in [23] and numerical simulations indicate that the map has no stable periodic orbits and possesses a smooth invariant density for $a < 2$ which develops a singularity $\sim |\varphi|^{-2/3}$ at $a = 2$[8]. In addition the results described below have also been found for the case of the "logistic map" $f(\varphi) = 2\varphi(2\pi - \varphi)/\pi$, $0 \leq \varphi \leq \pi$, $f(\varphi) = 2\pi - f(2\pi - \varphi)$, $\pi \leq \varphi \leq 2\pi$ which is obviously chaotic.

Although the Markov partition persists for non–hyperbolic coupled maps the expansivity is lost for very weak coupling. Hence even the weak coupling regime shows complex structures and it seems to be impossible to give a general description. But it has been stressed recently [15, 16, 17, 18] that there seems to exist some general feature among globally coupled non–hyperbolic maps on which I will focus in this section. They are related to an unusual system size dependence of the fluctuations of global quantities and have been termed "violation of the law of large numbers". It has been suggested that this effect is related to some hidden coherence in the coupled system. Partial explanations based on numerical evidence and stochastic arguments have been given. But there seems to exist no approach which explicitly refers to the non–hyperbolicity of the system. I will try to make some approach in this direction.

[8] The subsequent considerations demand for a generating partition. Its existence is guaranteed e.g in the case of expanding maps, $|(f^{(n)})'| > 1$ for some $n \in \mathbb{N}$.

Let me first briefly review the phenomenological results of numerical simulations (see also [15]). Owing to the global coupling of the maps the interaction is caused via a mean field

$$r_n \exp\left(i\Phi_n\right) := \frac{1}{L}\sum_{\nu} \exp\left(if(\varphi_n^{(\nu)})\right) \tag{23}$$

which is a global quantity. In terms of this field the dynamics of eq.(1) is written as

$$\varphi_{n+1}^{(\nu)} = F_n(f(\varphi_n^{(\nu)})), \quad F_n(\varphi) = \varphi + \epsilon r_n \sin(\Phi_n - \varphi) \tag{24}$$

where the interaction function $F_n$ depends on the other lattice sites via eq.(23). As long as the motion is chaotic the map (24) resembles the structure of a stochastically forced single map. A naive view of the law of large numbers suggests that the fluctuations of the mean field, $\Delta_{cc} := \langle r^2 \cos^2(\Phi)\rangle - \langle r\cos(\Phi)\rangle^2$, $\Delta_{ss} := \langle r^2 \sin^2(\Phi)\rangle - \langle r\sin(\Phi)\rangle^2$, will decrease with the system size as $\sim L^{-1}$. This behaviour is found in the hyperbolic case $a \ll 2$. But in non–hyperbolic situations strong deviations are observed. Fig.7 shows a few representative results for coupling strength $\epsilon = 0.2$ and < Fig.7
$a = 1.98$, $a = 2$. The fluctuations saturate at some finite value if the system size is increased beyond a critical value. The behaviour is more pronounced for the fluctuation $\Delta_{ss}$ and seems to increase by approaching $a = 2$ or increasing the coupling strength. Furthermore this ”violation of the law of large numbers” seems to occur for infinitesimal coupling strength in the non–hyperbolic case $a = 2$. But it is difficult to obtain conclusive results from the numerical simulations because of limitations in the system size. A much more refined view is obtained if the whole distribution function of the mean field is considered. Fig.8 shows some typical results which have been ob- < Fig.8
tained from a time series of length $N = 2.5 \times 10^6$ and a random initial condition. For small system size the distribution resembles a Gaussian shape. If the critical system size is approached the peak broadens and a double peak structure develops if the size is increased further. The spacing of the two maxima is responsible for the saturation of the moment $\Delta_{ss}$. It is furthermore instructive to look at the time evolution of the mean field (23). A finite part of the time series corresponding to parameter values used in Fig.8 are shown in Fig.9 . Whereas the evolution behaves random below the < Fig.9
critical system size an intermittent behaviour is clearly observed in the critical region. Beyond the critical length a hopping between states of positive and negative phases $\Phi$ at very low transition rate is detected. It may exceed the available time span of the numerical simulation. The numerical results resemble strongly the phenomena of symmetry breaking chaos transitions or crisis induced intermittency [24, 25, 26] which are well known in the context of low dimensional chaotic dynamics. An explanation in this direction will be developed in the sequel.

A partial analytical approach may be developed by studying the evolution of the full phase space distribution of the coordinates $\varphi^{(\nu)}$ which is governed by the Ruelle–Frobenius–Perron equation. This full description seems to be to complicated to handle with. But one may resort to a formulation introduced in ref.[27]. The global coupling enables us to reduce the full dynamics to the evaluation of a single site map (24) which depends on the mean field (23). Hence the motion can be analysed completely by considering the probability distribution that a lattice site assumes a value $\varphi$ at time $n$. It reads

$$\rho_n(\varphi) := \frac{1}{L}\sum_\nu \delta_{2\pi}(\varphi_n^{(\nu)} - \varphi) \tag{25}$$

where $\delta_{2\pi}$ denotes the $2\pi$ periodic extension of the $\delta$–distribution. An exact closed equation of motion for this density is easily written down. On one hand the density determines the mean field (23) via

$$r_n \exp(i\Phi_n) = \int \rho_n(\varphi)\exp(if(\varphi))\, d\varphi \quad . \tag{26}$$

On the other hand eq.(24) yields for the time evolution

$$\begin{aligned} \rho_{n+1}(\varphi) &= \frac{1}{L}\sum_\nu \delta_{2\pi}[(F_n\circ f)(\varphi^{(\nu)}) - \varphi] \\ &= \int \delta_{2\pi}[(F_n\circ f)(\psi) - \varphi]\rho_n(\psi)\, d\psi \quad . \end{aligned} \tag{27}$$

Eq.(27) resembles the structure of a Ruelle–Frobenius–Perron equation. But the dependence of the mean field, that means the map $F_n$, on the density itself turns the relation into a nonlinear evolution equation. On this level eq.(27) is an exact consequence of the full dynamical system (1) as long as the densities are of the form (25). For large system size these densities tend in a weak sense to smooth distributions for typical initial conditions. This suggestive observation has been checked numerically (cf. Fig.11). But also mathematical rigorous proofs are available in the hyperbolic case [28]. Hence eq.(27) yields the appropriate description of the dynamics in the limit of large system size if we consider sufficiently smooth densities[9]. It is much more easier to analyse than the full Ruelle–Frobenius–Perron equation of the coupled map lattice.

The long time behaviour is of central interest. Therefore let us concentrate on the discussion of stationary states and their stability. To this end let me briefly mention the symmetries of the coupled map lattice. The full system (1), (2), (3) is obviously invariant with respect to phase inversion, $\forall\nu, \varphi^{(\nu)} \to 2\pi - \varphi^{(\nu)}$. This property carries over to the mean field equation (27) which is invariant with respect to $\varphi \to 2\pi - \varphi$. Any solution is therefore either symmetric, $\rho_n^{(S)}(\varphi) = \rho_n^{(S)}(2\pi-\varphi)$ or has an inversion symmetric counterpart $\rho_n^{(-)}(\varphi) = \rho_n^{(+)}(2\pi - \varphi) \neq \rho_n^{(+)}(\varphi)$. In the former case the mean field is real, $\Phi_n^{(S)} = 0, \pi$, whereas in the latter case it takes complex conjugate values, $r_n^{(+)} = r_n^{(-)}$, $\Phi_n^{(+)} = -\Phi_n^{(-)}$. A real mean field does not destroy the inversion symmetry of the map (24) (cf. Fig.10) so that a symmetric solution stays symmetric <Fig.10

[9] It should be remarked that eq.(27) does not depend on $L$ explicitly. The system size enters only through the structure of the distribution. A smooth distribution corresponds to the limit $L \to \infty$

in the course of the dynamics.

The stationary densities $\rho_*(\varphi)$ are determined by

$$\rho_*(\varphi) = \int \delta_{2\pi}[(F_* \circ f)(\psi) - \varphi]\rho_*(\psi)\, d\psi \tag{28}$$

where the map $F_*$ depends on the field

$$r_* \exp(i\Phi_*) = \int \rho_*(\varphi) \exp(if(\varphi))\, d\varphi \quad . \tag{29}$$

Eq.(28) is formally a Ruelle–Frobenius–Perron equation for the one dimensional map $F_* \circ f$. The existence of smooth solutions and their computation is in general a difficult problem. At least for symmetric densities the results of [23] suggest that such a solution exists whereas numerical simulations indicate the existence of asymmetric solutions also for appropriate coupling strengths (cf. Fig.11). If one presupposes the existence of stationary smooth solutions their structure can be analysed in the non–hyperbolic case $a = 2$. In that case the point of slope zero, $\varphi = \pi$, introduces singularities into the density which can be computed by a perturbation expansion [29]. The case of symmetric densities is easy to handle. Here the point of slope zero is mapped onto the unstable fixed point (cf. Fig.10), a situation which is identical to the Smale complete logistic equation. The density $\rho_*^{(S)}$ develops a singularity $\rho_*^{(S)}(\varphi) \sim |\varphi|^{-2/3}$ at $\varphi = 0$ and is otherwise smooth. The case of asymmetric solutions is a little more difficult to handle (cf. appendix D). Here the point of slope zero has a non–trivial trajectory, $\omega_n = (F_* \circ f)^n(\pi)$. The density develops singularities at these image points , $\rho_*^{(\pm)} \sim |\varphi - \omega_n|^{-2/3}$ whose strengths decrease exponentially with $n$[10]. Hence the structure may become quite complicated. These results persist mainly for $a$ values near to the non–hyperbolic case, $2 - a \ll 1$. In that case the critical slope is non–zero but small and causes sharp peaks in the density in contrast to actual singularities. The symmetric solution is smooth and has a sharp maximum at the unstable fixed point whereas the asymmetric density develops peaks on the orbit of the critical point. For small coupling strength $\epsilon$ this orbit yields a sequence which departs exponentially from the unstable fixed point (cf. Fig.11 for comparison with numerical simulations). Finally it should be mentioned that the described enhancement of the density reflects a partial synchronization among lattice sites in the full system. This effect is attributed to the non–hyperbolicity of the single site map.

Let us give some qualitative estimates on the stability of the stationary density. As long as linear stability analysis is valid it is determined by the eigenvalue problem which emerges from the linearization of the evolution equation (27). It reads

$$\lambda h_\lambda(\varphi) \quad = \quad \int \delta[\varphi - (F_* \circ f)(\psi)] h_\lambda(\psi)\, d\psi$$

[10] The result is rigorous if the orbit terminates at some unstable periodic orbit [23].

$$-\epsilon \mathrm{Im}\left(\int \delta'[\varphi - (F_* \circ f)(\psi)]\rho_*(\psi)\exp[-if(\psi)]\,d\psi \right.$$
$$\left. \cdot \int \exp[if(\psi)]h_\lambda(\psi)\,d\psi\right) \quad . \tag{30}$$

In order that this expression makes sense the density $\rho_*$ should be differentiable. For that reason let us restrict the subsequent discussion to the case of symmetric densities and $a < 2$. Integrating the eigenvalue equation (30) with respect to $\varphi$ one obtains immediately that eigenvalues $\lambda \neq 1$ are necessarily related to eigenfunctions with vanishing average value, $\int h_\lambda(\varphi)\,d\varphi = 0$. Furthermore there exists an eigenvalue $\lambda = 1$ which is independent of the coupling strength $\epsilon$. This "Goldstone mode" is related to a continuous symmetry of the fixed point equation (28)[11]. For small coupling eq.(30) yields a perturbation of the Ruelle–Frobenius–Perron equation of the symmetric map $F_* \circ f$. The unperturbed eigenvalue problem possesses by presupposition (cf. the remarks at the beginning of this section) one eigenvalue $\lambda = 1$. The remaining part of the spectrum has modulus smaller than 1 and is separated from the largest eigenvalue by a finite gap. This gap shrinks to zero if $a$ approaches 2. The largest eigenvalue $\lambda = 1$ persists if we turn on a small coupling $\epsilon > 0$. We need a finite value in order that additional eigenvalues cross the line $\mathrm{Re}(\lambda) = 1$ and induce the instability of the symmetric density. The strength of this critical coupling is supposed to tend to zero in the limit $a \uparrow 2$ because the gap vanishes in that limit. Hence in the non–hyperbolic case even an infinitesimal coupling $\epsilon$ may cause the instability of the symmetric density because the leading eigenvalue $\lambda = 1$ is degenerated with a continuous part of the spectrum [29, 30]. Summarizing this reasoning one may expect that the symmetric density, $\Phi_* = 0$, is stable up to a critical coupling strength. This critical value tends to zero in the limit $a \uparrow 2$.

Further analysis is possible if we look at numerical solutions of the mean field equation (27). This analysis captures also dynamical properties of the system. To implement the numerical algorithm a sufficiently fine partition of the phase space $[0, 2\pi]$ in intervals of equal size is considered. The densities $\rho_n$ are approximated by step functions on this partition. Eq.(27) then turns into a matrix equation which can be iterated easily. For our purpose a number of $\sim 10^4$ boxes is sufficient to give accurate results for the densities too. First of all it has been checked that the mean field equation yields the correct description in the limit of large system size. As a typical example Fig.11 displays the stationary densities for to parameter combinations $\epsilon = 0.05$, $a = 1.99, 1.996$ in comparison with the direct numerical simulation of a huge coupled map lattice $L = 5 \times 10^5$. These densities have been obtained from the time evolution of the map lattice (1) and the mean field equation (27) which tend to a time independent state for the chosen parameter values after 100 time steps. The first set of parameters yields a symmetric density with a sharp maximum at the unstable fixed <Fig.11 point whereas for the second combination of parameters a non–symmetric density is observed. Its maxima escape from the unstable fixed point with an exponential rate as predicted by the analytical analysis. The mean field result and the direct simulation coincide within the statistical errors which are caused by finite size effects. Hence the system shows the two types of stationary solutions described above. Beyond the stationary properties the mean field approach reproduces the dynamical features also. Fig.11c shows a typical time evolution of the phase of the mean field $\Phi_n$ obtained from a random initial condition, that means a constant initial density. Again the mean field description and the simulation coincide. Let us now study the stability of the symmetric solution in the weak coupling case. Because analytical approaches seem to be difficult to apply the results of a numerical integration of the mean field equation (27) are presented. For sufficiently small coupling $\epsilon$ one finds a stable symmetric stationary density for $a < 2$. If the coupling is increased this solution becomes unstable. Fig.12 shows the time evolution of the phase of the mean field, $\Phi_n$, that emerges from <Fig.12 a slightly asymmetric initial condition. If we disregard a short oscillatory transient then an exponential relaxation towards the symmetric solution is observed below a critical coupling strength for $a < 2$. An exponential relaxation towards an asymmetric stationary state is observed if we cross the critical coupling strength. The critical value decreases if one approaches $a = 2$. In the non–hyperbolic case $a = 2$ for every coupling strength only a stable asymmetric state (cf. Fig.12d) was found. The hole scenario is consistent with the qualitative linear stability analysis given above.

[11] Denoting by $\rho_*(\varphi, \epsilon)$ a (non–normalized) solution of eq.(28) then $\alpha\rho_*(\varphi, \epsilon) = \rho_*(\varphi, \epsilon/\alpha)$, $\alpha > 0$ holds.

The numerical treatment of the mean field equation enables us to investigate regions of larger coupling strength also which seem to be inaccessible by means of an analytical approach. The time evolution of the system may become quite complicated. Typically I have observed a transition from stationary to oscillatory and time chaotic behaviour. As a typical example Fig.13 contains the time series of the phase of the <Fig.13 mean field for increasing coupling strength and two parameter values $a = 1.99, 1.95$. Although this scenario which resembles on a phenomenological level the transition to chaos in low dimensional dynamical systems [3] has been observed for almost all combinations of parameter values it is not clear to which extent it depends on the details of the system under consideration.

# 5 Conclusion

For a certain class of globally coupled map lattices the explicit construction of a Markov partition has lead us to a detailed insight into the dynamics. For coupled shift maps, $a = 0$, all invariant sets have been discovered. Beyond the weak coupling regime which is governed by a space–time–mixing stationary state a global synchronization sets in. It is accompanied by transients whose length increase exponentially with the system size. The scenario extends to perturbations of the shift map, $a > 0$.

But then there might occur a gap between the onset of synchronization at $\epsilon_c^s$ and the loss of space–time–mixing, that means the loss of local expansivity, at $\epsilon_c^e$. It is however difficult to observe such phenomena in numerical simulations because of the tremendous transients. If the coupling is increased to very large values, $\epsilon \gg \epsilon_c^m$, then the synchronized state undergoes further bifurcations and may be destroyed again. But these effects seem to depend strongly on the nature of the coupling in contrast to the above mentioned properties.

A different behaviour is observed for strongly non–hyperbolic coupled maps, $a \lesssim 2$. For small coupling a symmetry breaking is observed in the thermodynamic limit of infinite system size. This bifurcation is triggered by the non–hyperbolic points and can be understood qualitatively in terms of a mean field transfer operator. Fluctuations which are introduced by finite size effects cause jumps between the different asymmetric states. They are responsible for the bimodal structure of the distribution of the mean field and the anomalous behaviour of the mean square deviations. If the size of the system is increased the time scale of the intermittent hopping may exceed the observation time so that the computed mean square deviation drops again. But then secondary bifurcations to time dependent stationary states lead to a lower bound for the mean square deviations. Whether the symmetry breaking of the inversion symmetry is a necessary condition for the subsequent bifurcations like in low dimensional systems [31] is an unsolved problem. Map lattices without inversion symmetry like the frequently discussed coupled logistic maps are believed to undergo these "secondary" bifurcations. This scenario will explain the so called "violation of the law of large numbers". Results will be published elsewhere.

Depending on the non–hyperbolicity of the single site map the breaking of the inversion symmetry precedes the appearance of a synchronized state. For intermediate values of the parameter $a$ both bifurcations may interact. The consequences of such higher order codimension bifurcation is at the moment not clear. Furthermore it is an open question whether it influences the dynamics on a time scale that is observable in numerical simulations of large coupled map lattices.

The model system proposed in this publication constitutes an example which can be handled to a great extent by analytical methods. It seems therefore promising to construct explicitly the two dimensional spin Hamiltonian and to analyse how the dynamical synchronization is related to phase transitions in the corresponding spin lattice. Such a relation might be helpful to understand pattern formation in more complicated and realistic systems. Work in this direction is in progress.

# Acknowledgement

The author is indebted to the Deutsche Forschungsgemeinschaft for financial support. This work was performed within a program of the Sonderforschungsbereich 185

Darmstadt–Frankfurt, FRG.

# Note added in proof

After the submission of the manuscript I got knowledge of ref.[32] which treats a related subject. I also thank the referee for pointing to this publication.

# Appendix A

The determinant of the expression (5) reads

$$\det(D\tilde{T}(\underline{x}) = \prod_{\nu=0}^{L-1} f'(x^{(\nu)}) \det(\underline{\underline{A}}) \tag{31}$$

where the matrix on the right hand side is given by

$$A_{\nu\mu} = \delta_{\nu\mu} - \frac{\epsilon}{L}\sum_{\rho}\gamma_{\nu\rho}\delta_{\nu\mu} + \frac{\epsilon}{L}\gamma_{\nu\mu} \quad . \tag{32}$$

For simplicity the abbreviation

$$\gamma_{\nu\mu} := g'\left(f(x^{(\mu)}) - f(x^{(\nu)})\right) = \cos\left((f(x^{(\mu)}) - f(x^{(\nu)})\right) \tag{33}$$

has been used. We calculate the matrix of the expression (32) via its eigenvalue equation. It is written as

$$\Lambda_\nu v^{(\nu)} = \frac{\epsilon}{L}\left(c_\nu \sum_\mu c_\mu v^{(\mu)} + s_\nu \sum_\mu s_\mu v^{(\mu)}\right) \tag{34}$$

where use has been made of the abbreviations

$$\begin{aligned} \Lambda_\nu &:= \lambda - 1 + \frac{\epsilon}{L}\sum_\mu \gamma_{\nu\mu} \\ c_\nu + is^{(\nu)} &:= \exp(if(x^{(\nu)})) \end{aligned} \tag{35}$$

and the trigonometric identity

$$\gamma_{\nu\mu} = c_\nu c_\mu + s_\nu s_\mu \quad . \tag{36}$$

$\lambda$ denotes the eigenvalue and $v^{(\nu)}$ the components of the eigenvector. Introducing the two real quantities $\alpha, \beta$

$$\alpha + i\beta := \sum_\mu (c_\mu + is_\mu) v^{(\mu)} \tag{37}$$

the eigenvalue equation (34) reduces to

$$\begin{aligned}\alpha &= \frac{\epsilon}{L}\sum_\nu c_\nu \frac{c_\nu\alpha + s_\nu\beta}{\Lambda_\nu} \\ \beta &= \frac{\epsilon}{L}\sum_\nu s_\nu \frac{c_\nu\alpha + s_\nu\beta}{\Lambda_\nu} \quad .\end{aligned} \tag{38}$$

The condition for a non–trivial solution of this homogeneous linear equation yields a polynomial of degree $L$ in the eigenvalue $\lambda$

$$\prod_\nu \Lambda_\nu - \frac{\epsilon}{L}\sum_\nu \prod_{\rho(\neq\nu)} \Lambda_\rho + \frac{1}{2}\left(\frac{\epsilon}{L}\right)^2 \sum_{\nu\neq\rho}(1-\gamma_{\nu\rho}^2) \prod_{\mu(\neq\nu,\rho)} \Lambda_\mu = 0 \quad . \tag{39}$$

Eq.(39) is the characteristic polynomial of the matrix (32). Hence its value at $\lambda = 0$ yields the determinant of the matrix. Combining this expression with eq.(31) leads to the result (6).

# Appendix B

Presuppose that the map $\tilde{T} : \mathbb{R}^L \to \mathbb{R}^L$ is sufficiently smooth and locally expansive, that means that eq.(9) holds. Consider an arbitrary phase space point $\underline{x}$ and let

$$U_\delta(\underline{x}) := \{\underline{y} \in \mathbb{R}^L \mid \|\underline{y} - \underline{x}\| < \delta\} \tag{40}$$

denote a $\delta$ neighbourhood of this point. The image $\tilde{T}(U_\delta(\underline{x}))$ contains a $\delta'$ neighbourhood of $\underline{x}' = \tilde{T}(\underline{x})$, $U_{\delta'}(\underline{x}')$. The largest neighbourhood is obtained if we choose the distance between the point $\underline{x}'$ and the boundary of the image as $\delta'$. It will be shown that $\delta' > \delta$ holds. Denote by $\underline{y}'$ a common point of the boundaries of $U_{\delta'}(\underline{x}')$ and $\tilde{T}(U_\delta(\underline{x}))$. Let $\underline{s}'(t) = t\underline{x}' + (1-t)\underline{y}'$, $t \in [0,1]$ be the straight line which connects these points. It possesses a unique preimage $\underline{s}(t)$ with $\underline{s}(0) = \underline{x}$ and $\underline{s}(1) = \underline{y}$. By construction the final point $\underline{y}$ is contained in the boundary of the set (40). We now have

$$\delta' = \|\underline{y}' - \underline{x}'\| = \int_0^1 \left\| \frac{d\tilde{T}(\underline{s})}{dt} \right\| dt \geq \int_0^1 \left\| D\tilde{T}(\underline{s}(t)) \frac{d\underline{s}}{dt} \right\| dt \geq c\delta \quad . \tag{41}$$

The last inequality follows from the condition of local expansivity (cf. eq(9)) and the fact that the path $\underline{s}(t)$ has at least the length $\delta$.

Denote by $\tilde{T}$ the extension of the coupled map lattice and by $\tilde{\Lambda}$ the $2\pi$ periodic continuation of the invariant set on which the map is locally expansive. Suppose $\tilde{\Lambda}$ contains some neighbourhood. We will show that it coincides with the whole phase space. To this end suppose that for some point $\underline{x} \in \tilde{\Lambda}$ there exists a $\delta > 0$ so that $U_\delta(\underline{x}) \subseteq \tilde{\Lambda}$ holds. Because of the invariance of $\tilde{\Lambda}$ every image of the neighbourhood is contained in the invariant set, $\tilde{T}^n(U_\delta(\underline{x})) \subseteq \tilde{\Lambda}$. Hence we can apply the global expansion property which states that the set $\tilde{T}^n(U_\delta(\underline{x}))$ contains a neighbourhood with a diameter $c^n\delta$. Choosing $n$ large enough it can be managed this set contains a full hypercube $[0, 2\pi]^L$. Hence the invariant set $\tilde{\Lambda}$ contains a full hypercube. But then this $2\pi$ periodic set coincides with the whole phase space.

# Appendix C

Let $A$ denote a neighbourhood of the attractor. Then eqs.(21) and (22) immediately lead to

$$\sum_{k=0}^{n} P_k(A) = \lambda(T^{-n}(A)) \tag{42}$$

where $\lambda$ denotes the normalized measure which yields the average in eq.(22). A natural choice for this measure is the normalized Lebesgue measure which corresponds to a uniform distribution of initial conditions. Let us presuppose that the quantity (42) tends to 1 in the limit $n \to \infty$. It means that the system has only one attracting set and that all repellers have (Lebesgue) measure zero. From the normalization of the measure and the identity $T^{-n}(S^L/A) = S^L/T^{-n}(A)$ one obtains

$$\sum_{k=n+1}^{\infty} P_k(A) = \lambda(T^{-n}(S^L/A)) \quad . \tag{43}$$

The conventional definition of the escape rate from a repelling set $\Lambda$ is given by [22]

$$\sigma = -\lim_{U_\epsilon \downarrow \Lambda} \lim_{n\to\infty} \frac{1}{n} \ln \lambda(T^{-n}(U_\epsilon) \tag{44}$$

where $U_\epsilon$ denotes an $\epsilon$ neighbourhood of the repelling set. It is assumed that the escape rate does not depend on the choice of the (sufficiently small) neighbourhood so that one limit can be suppressed. The set $S^L/A$ constitutes a suitable neighbourhood. Then eq.(44) yields the asymptotic behaviour

$$\lambda(T^{-n}(S^L/A)) \simeq C_1 \exp(-n\sigma) \quad . \tag{45}$$

Eqs.(43) and (45) result in the asymptotic relation

$$P_n(A) \simeq C_2 \exp(-n\sigma) \quad . \tag{46}$$

# Appendix D

Eq.(28) can be viewed as a Ruelle–Frobenius–Perron equation for the asymmetric map $F_* \circ f$. It may be solved by iterating a smooth initial density $h_{n=0}(\varphi)$ until a stationary state is reached [29]

$$h_{n+1}(\varphi) = \int \delta_{2\pi}[(F_{\Phi_*} \circ f)(\psi) - \varphi] h_n(\psi)\, d\psi \quad . \tag{47}$$

For later use the explicit dependence of the map $F_*$ on the phase of the mean field, $\Phi_*$, has been indicated. Evaluating the integral the relation can be written as

$$h_{n+1}(\varphi) = \frac{h_n[(F_{\Phi_*} \circ f)^{-1}(\varphi)]}{(F_{\Phi_*} \circ f)'[(F_{\Phi_*} \circ f)^{-1}(\varphi)]} + \frac{h_n[2\pi - (F_{-\Phi_*} \circ f)^{-1}(2\pi - \varphi)]}{(F_{-\Phi_*} \circ f)'[(F_{-\Phi_*} \circ f)^{-1}(2\pi - \varphi)]} \tag{48}$$

where the the function $f$ is restricted to the interval $[0, \pi]$. If the initial function $h_{n=0}(\varphi)$, is smooth the first iterate develops singularities at the critical points of the denominator. They are determined by

$$\begin{aligned} (F_{\Phi_*} \circ f)^{-1}(\varphi) \uparrow \pi \quad &\Leftrightarrow \quad \varphi \uparrow \omega_1 \\ (F_{-\Phi_*} \circ f)^{-1}(2\pi - \varphi) \uparrow \pi \quad &\Leftrightarrow \quad x \downarrow \omega_1 \quad . \end{aligned} \tag{49}$$

where $\omega_1 := \epsilon r_* \sin(\Phi_*)$ denotes the first image of the critical point. The strength of the singularity can be evaluated easily by considering neighbourhood of this point.

*i)* $\varphi \downarrow \omega_1$: Straightforward Taylor series expansion yields to the leading order

$$\begin{aligned} (F_{\Phi_*} \circ f)^{-1}(\varphi) &= \frac{\varphi - \omega_1}{4F'_{\Phi_*}(0)} + O(2) \\ (F_{\Phi_*} \circ f)'[(F_{\Phi_*} \circ f)^{-1}(\varphi)] &= 4F'_{\Phi_*}(0) + O(1) \\ (F_{-\Phi_*} \circ f)^{-1}(2\pi - \varphi) &= \pi - \left(3\frac{\varphi - \omega_1}{F'_{-\Phi_*}(0)}\right)^{1/3} + O(2/3) \\ (F_{-\Phi_*} \circ f)'[(F_{-\Phi_*} \circ f)^{-1}(2\pi - \varphi)] &= F'_{-\Phi_*}(0)\left(3\frac{\varphi - \omega_1}{F'_{-\Phi_*}(0)}\right)^{2/3} + O(1) \end{aligned} \tag{50}$$

where $O(n)$ denotes contributions of order $(\varphi - \omega_1)^n$. Then eq.(48) reads

$$h_{n+1}(\varphi) \simeq \frac{h_n[(\varphi - \omega_1)/(4F'_{\Phi_*}(0))]}{4F'_{\Phi_*}(0)} + \frac{h_n(\pi)}{F'_{-\Phi_*}(0)[3(\varphi - \omega_1)/F'_{-\Phi_*}(0)]^{2/3}} \quad . \tag{51}$$

*ii)* $\varphi \uparrow \omega_1$: An analogous expansion leads to

$$\begin{aligned} (F_{\Phi_*} \circ f)^{-1}(\varphi) &= \pi - \left(3\frac{\omega_1 - \varphi}{F'_{\Phi_*}(0)}\right)^{1/3} + O(2/3) \\ (F_{\Phi_*} \circ f)'[(F_{\Phi_*} \circ f)^{-1}(\varphi)] &= F'_{\Phi_*}(0)\left(3\frac{\omega_1 - \varphi}{F'_{\Phi_*}(0)}\right)^{2/3} + O(1) \\ (F_{-\Phi_*} \circ f)^{-1}(2\pi - \varphi) &= \frac{\omega_1 - \varphi}{4F'_{-\Phi_*}(0)} + O(2) \\ (F_{-\Phi_*} \circ f)'[(F_{-\Phi_*} \circ f)^{-1}(2\pi - \varphi)] &= 4F'_{-\Phi_*}(0) + O(1) \end{aligned} \tag{52}$$

and

$$h_{n+1}(\varphi) \simeq \frac{h_n(\pi)}{F'_{\Phi_*}(0)[3(\omega_1 - \varphi)/F'_{\Phi_*}(0)]^{2/3}} + \frac{h_n[2\pi - (\omega_1 - \varphi)/(4F'_{-\Phi_*}(0))]}{4F'_{-\Phi_*}(0)} \quad . \tag{53}$$

Eqs.(51) and (53) clearly show that the first iterate develops a singularity $|\varphi - \omega_1|^{-2/3}$. On further iteration this singularity is carried along the trajectory of $\omega_1$. With $\omega_2 = (F_* \circ f)(\omega_1)$ eq.(47) yields in the vicinity of the image point

$$\begin{aligned} h_{n+1}(\varphi) &\sim \int \delta_{2\pi}[(F_* \circ f)(\psi) - \varphi]|\psi - \omega_1|^{-2/3}\, d\psi \\ &\sim |(F_* \circ f)'(\omega_1)|^{-1/3}|\varphi - \omega_2|^{-2/3} \quad . \end{aligned} \tag{54}$$

Hence the strength of the singularity decreases by a factor $|(F_* \circ f)'(\omega_1)|^{-1/3} \sim 4^{-1/3}$ along the trajectory $\omega_n$.

# Figure captions

Fig.1 Diagrammatic view of the Markov partition $U_{\sigma^{(0}\sigma^{(1)}}$ for two coupled maps ($L = 2$) in the extended phase space. The dotted lined indicates the partition for the uncoupled case $\epsilon = 0$, and the broken line the partition for finite coupling $\epsilon > 0$.

Fig.2 Diagrammatic view of the map $f^{(-)}(x^{(-)}) = 2x^{(-)} - \epsilon \sin(2x^{(-)})\ |\mathrm{mod}\, 2\pi$ for $\epsilon > \epsilon_c^s$. The box indicates the region which contains the repelling Cantor set.

Fig.3 Time series of the amplitude of the mean field for different coupling strengths and system sizes. (a) $\epsilon = 0.65$, $L = 15$, (b) $\epsilon = 0.65$, $L = 21$, (c) $\epsilon = 0.8$, $L = 15$, and (d) $\epsilon = 0.8$, $L = 21$.

Fig.4 Distribution of the relaxation times for $\epsilon = 0.8$ and (a) $L = 12$, (b) $L = 15$. The broken line indicates the least square fit. It is shifted by a factor two for clarity.

Fig.5 Dependence of the inverse of the mean relaxation time on the system length for several coupling strengths.

Fig.6 Dependence of the slopes in Fig.5 on the coupling strength $\epsilon - \epsilon_c^s$. The full line indicates the least square fit.

Fig.7 Mean square deviations $\Delta_{cc}$ (————) and $\Delta_{ss}$ (– – – –) in dependence on the system size for two values of the nonlinearity parameter $a = 1.98$ ($\times$) and $a = 2$ ($\square$).

Fig.8 Distribution function of the mean field for $a = 1.98$, $\epsilon = 0.2$ and several values of the system size, (a) $L = 320$, (b) $L = 640$, and (c) $L = 1280$.

Fig.9 Time evolution of the mean field corresponding to the parameter values chosen in Fig.8, $a = 1.98$, $\epsilon = 0.2$ and (a) $L = 320$, (b) $L = 640$, (c) $L = 1280$. The upper chain of dots indicates the real part $r_n \cos(\Phi_n)$ and the lower chain the imaginary part $r_n \sin(\Phi_n)$.

Fig.10 Diagrammatic view of the stationary mean field map $F_* \circ f$ for (a) $\Phi_* = 0$ and (b) $\Phi_* > 0$. The broken line indicates the orbit of the critical point.

Fig.11 (a) Symmetric distribution of the phase space coordinates for $a = 1.99$ and $\epsilon = 0.05$. The full line indicates the result of a simulation of a map lattice of size $L = 5 \times 10^5$ whereas the broken line shows the mean field distribution. The latter curve is shifted by a factor 1/3 for clarity because both curves would coincide otherwise. $10^3$ boxes (simulation) respectively $10^4$ boxes (mean field) are used to figure the distribution.

(b) Asymmetric distribution for $a = 1.996$, $\epsilon = 0.05$ on a double logarithmic scale. The lower broken line indicates the mean field result whereas the upper broken line is shifted by a factor 3. $10^4$ boxes (simulation) respectively $4 \times 10^4$ boxes (mean field) are used to figure the distribution.
(c) Time dependence of the phase of the mean field for parameter values used in Fig.11b, $a = 1.996$, $\epsilon = 0.05$. The full line indicates the mean field result whereas the broken line the simulation of the map lattice ($L = 5 \times 10^5$).

Fig.12 Time evolution of the phase of the mean field for several values of the coupling strength and nonlinearity parameter, (a) $a = 1.97$, (b) $a = 1.98$, (c) $a = 1.99$, and (d) $a = 2$.

Fig.13 Time evolution of the phase of the mean field for larger coupling strength.
(a) $a = 1.99$. The coupling strength is chosen as (from bottom to top) $\epsilon = 0.1$, 0.125, 0.15, 0.175, 0.2. The curves are shifted against each other by an offset of 0.1.
(b) $a = 1.95$. The coupling strength is chosen as (from bottom to top) $\epsilon = 0.4$, 0.5, 0.6, 0.7, 0.8. The curves are shifted against each other by an offset of 0.4.

Fig.1

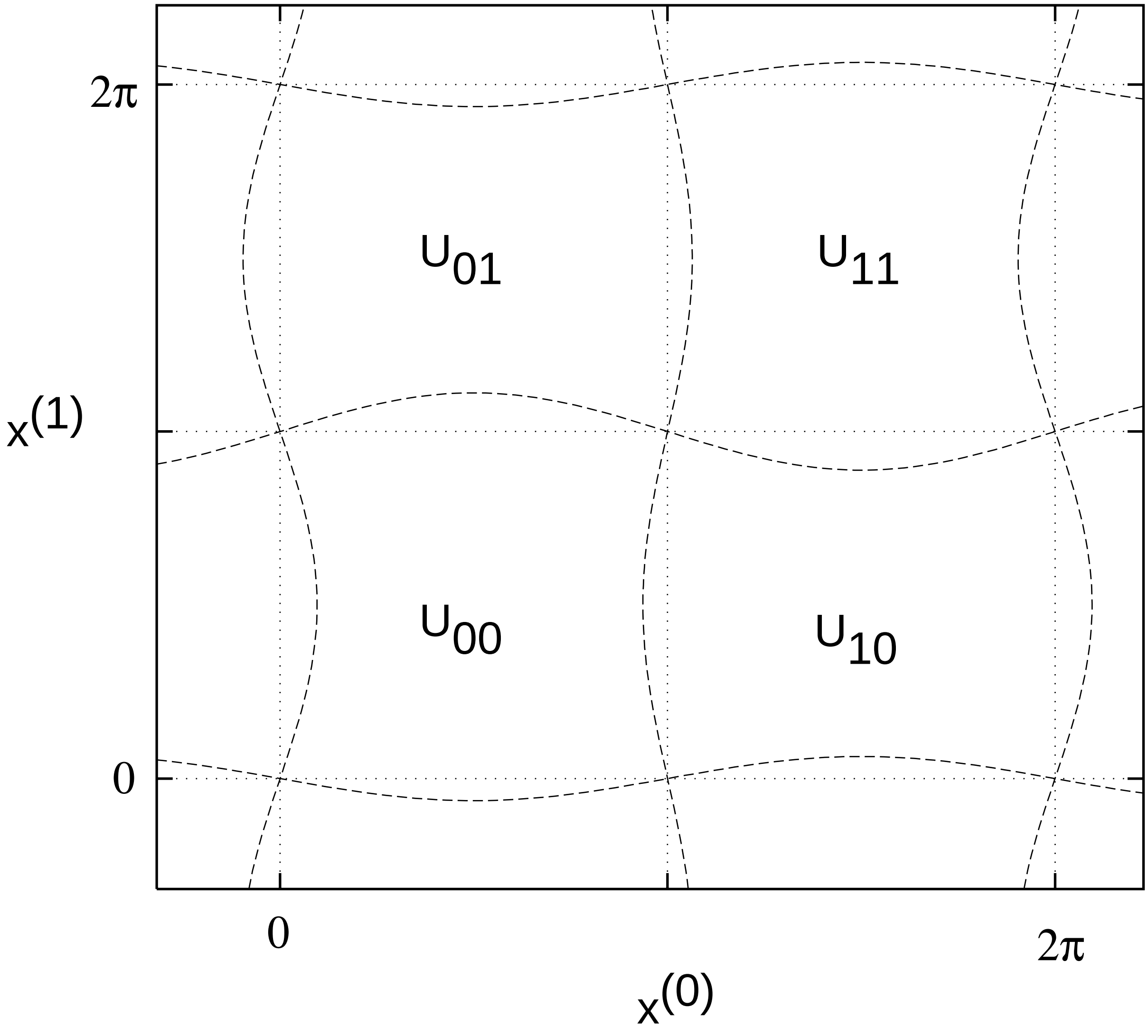

Fig.2

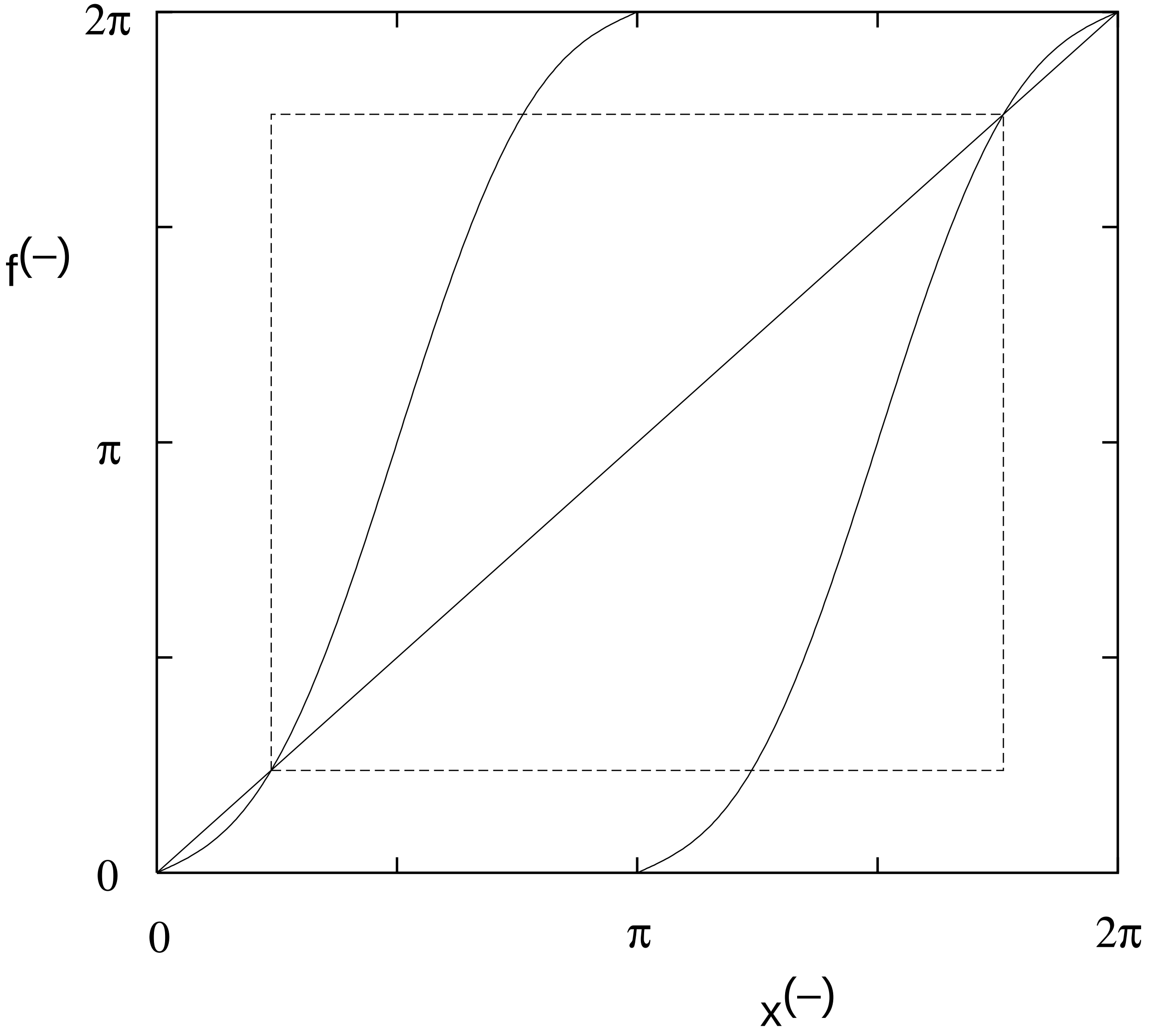

Fig.3a

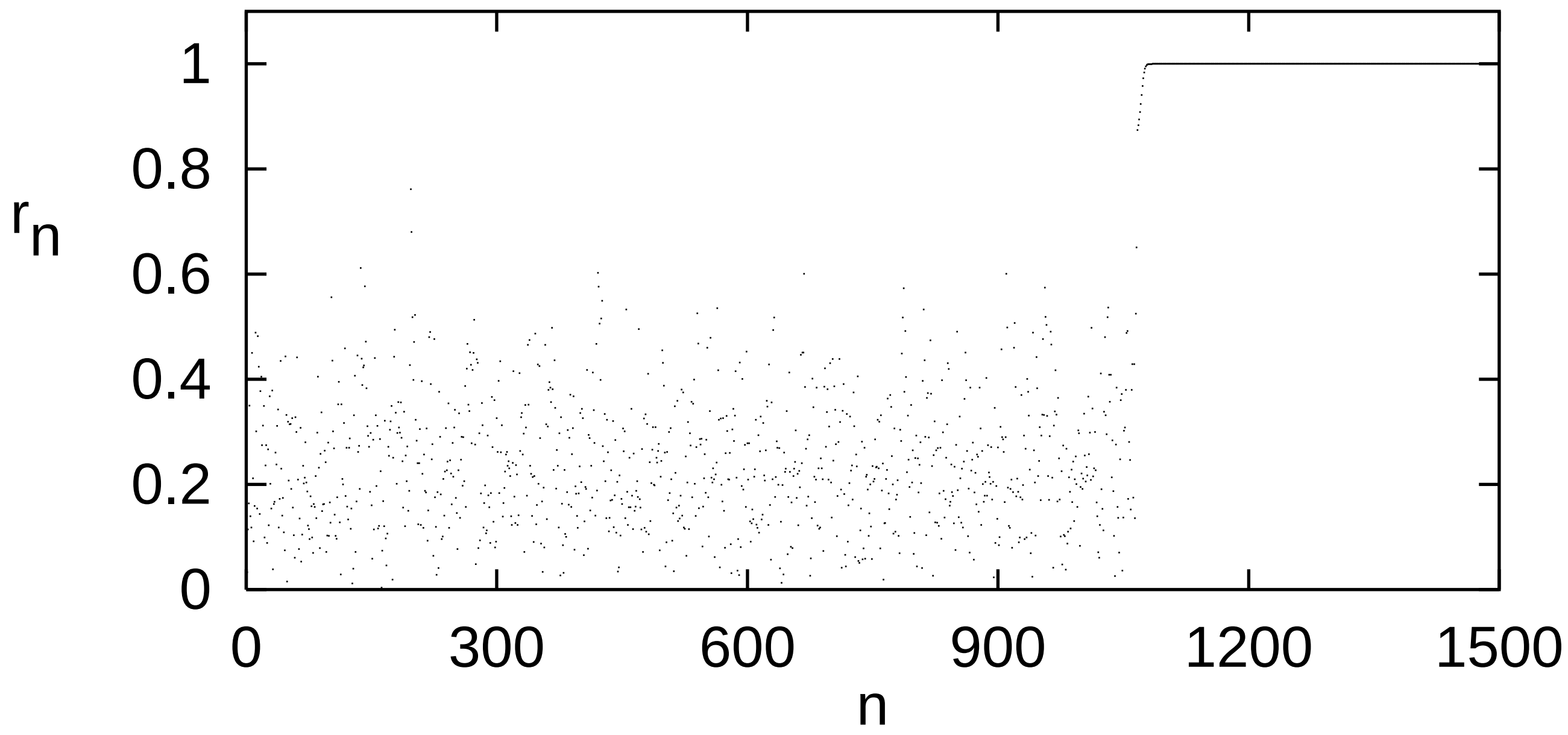

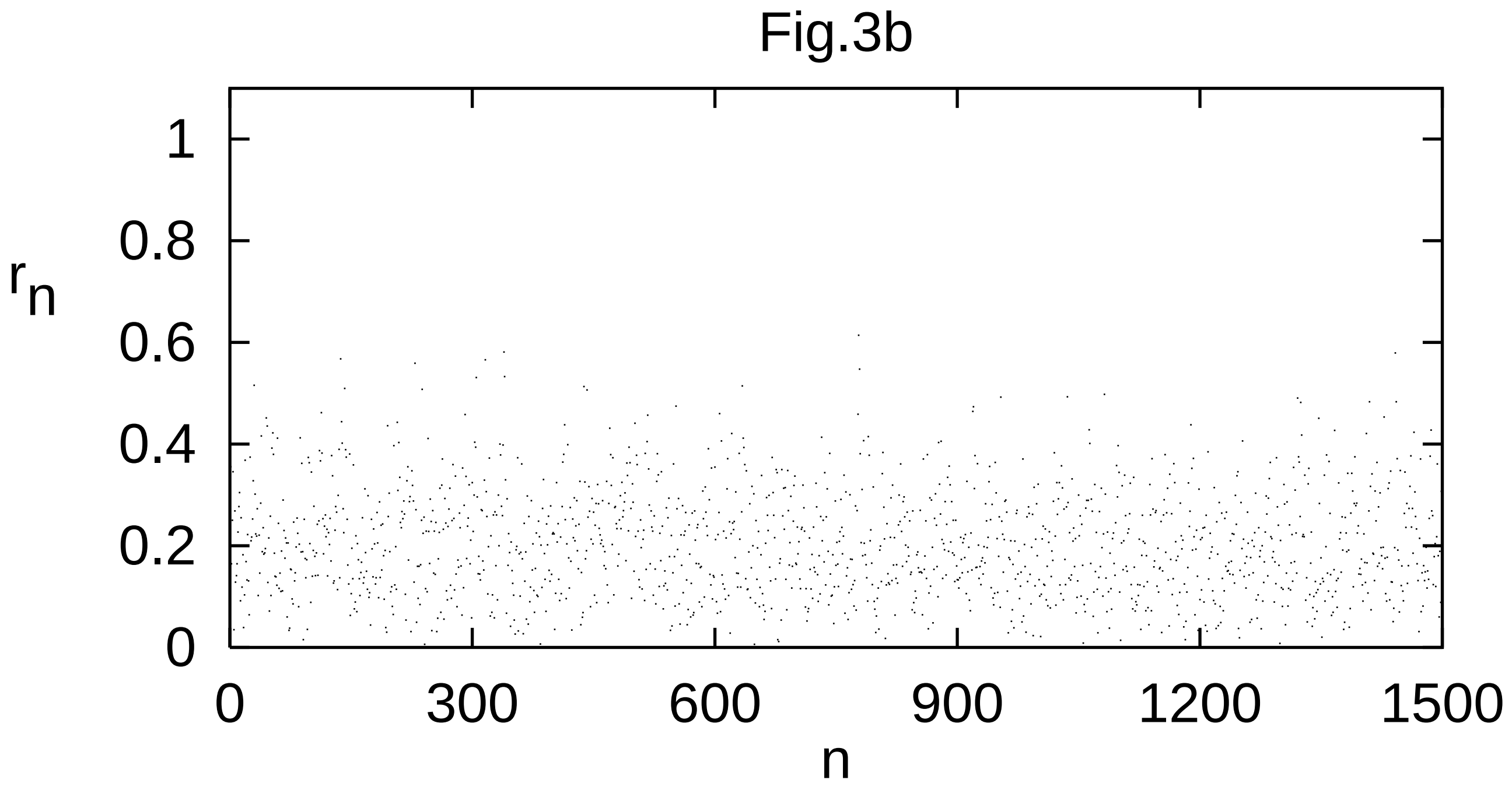

Fig.3c

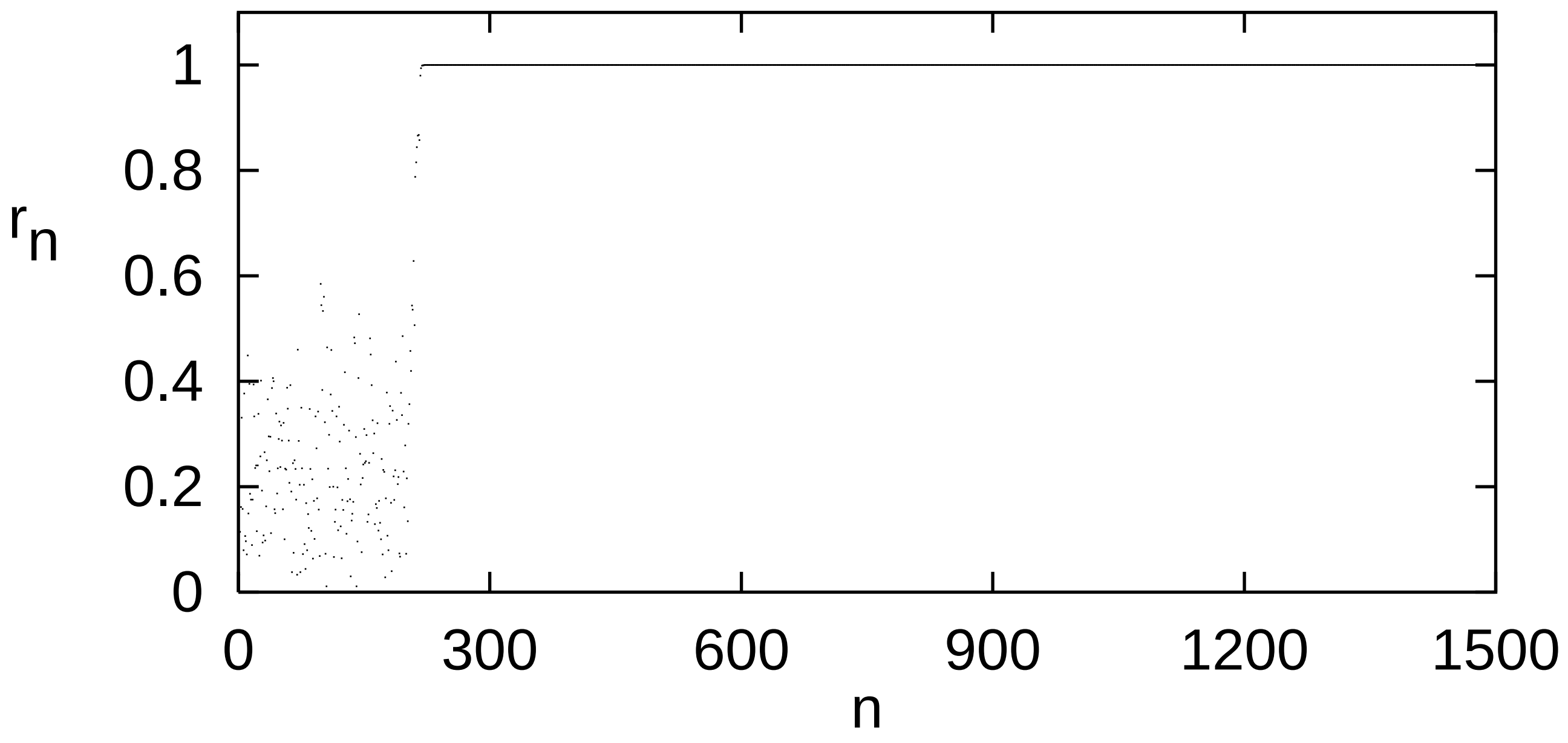

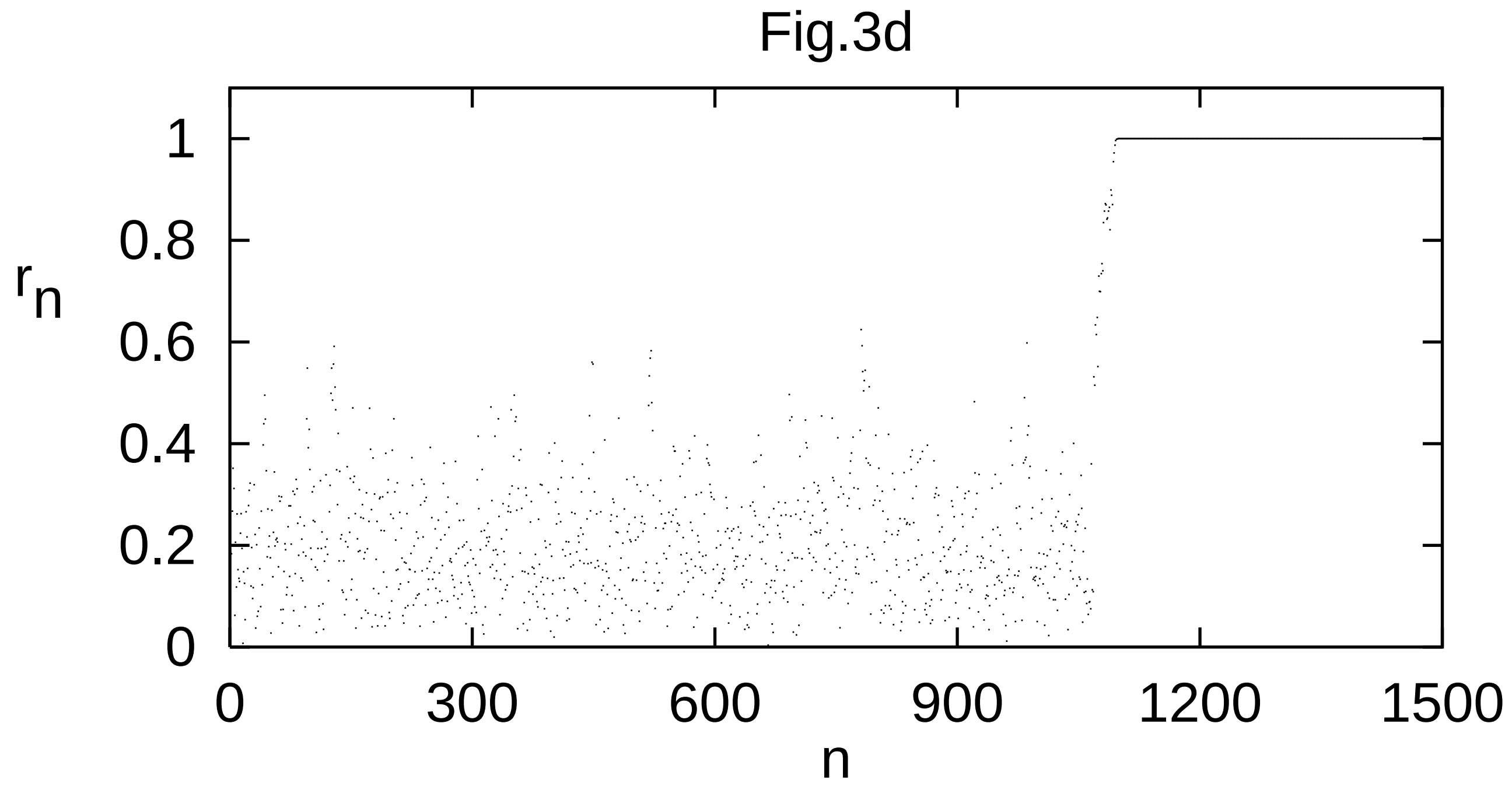

Fig.4a

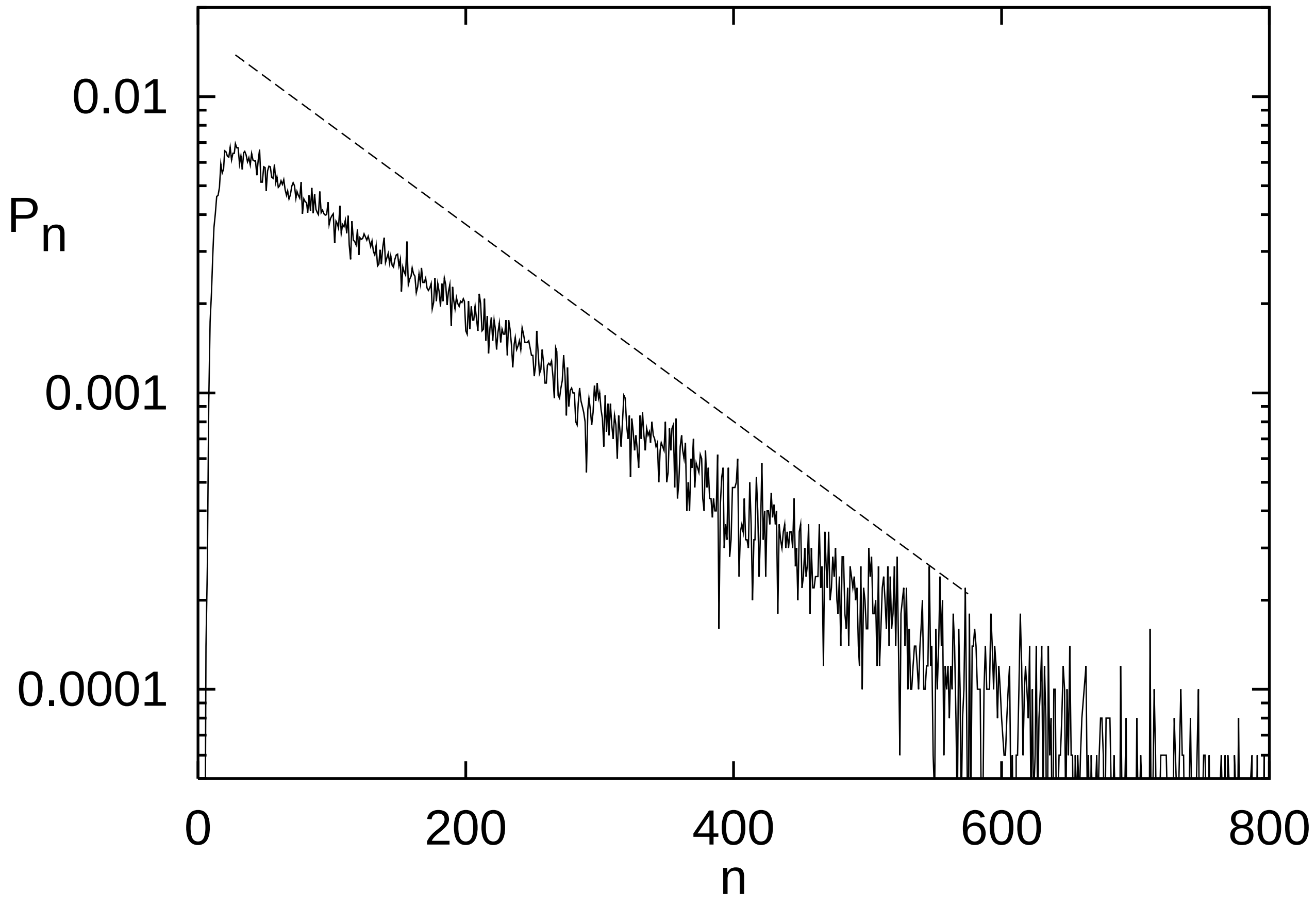

Fig.4b

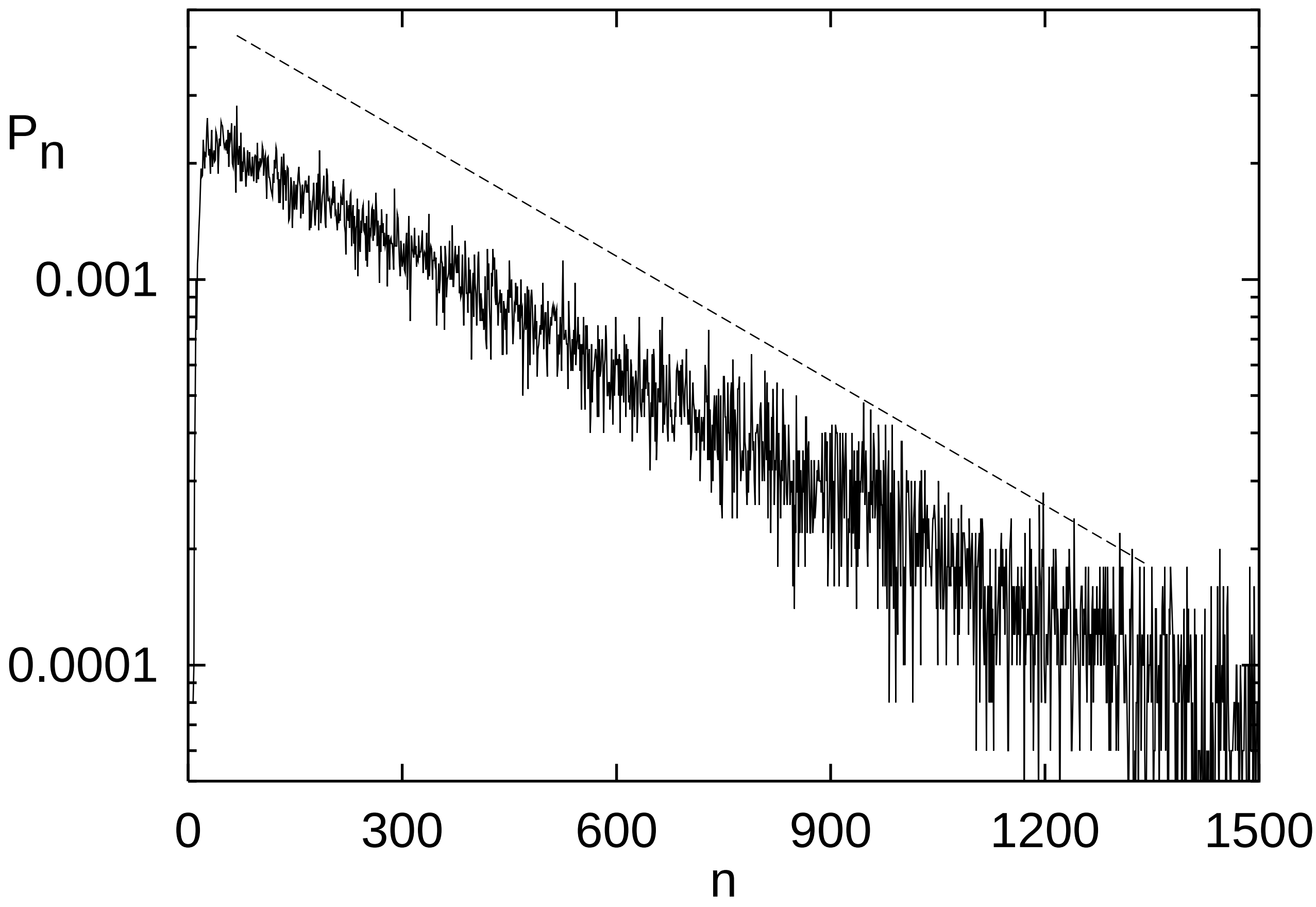

Fig.5

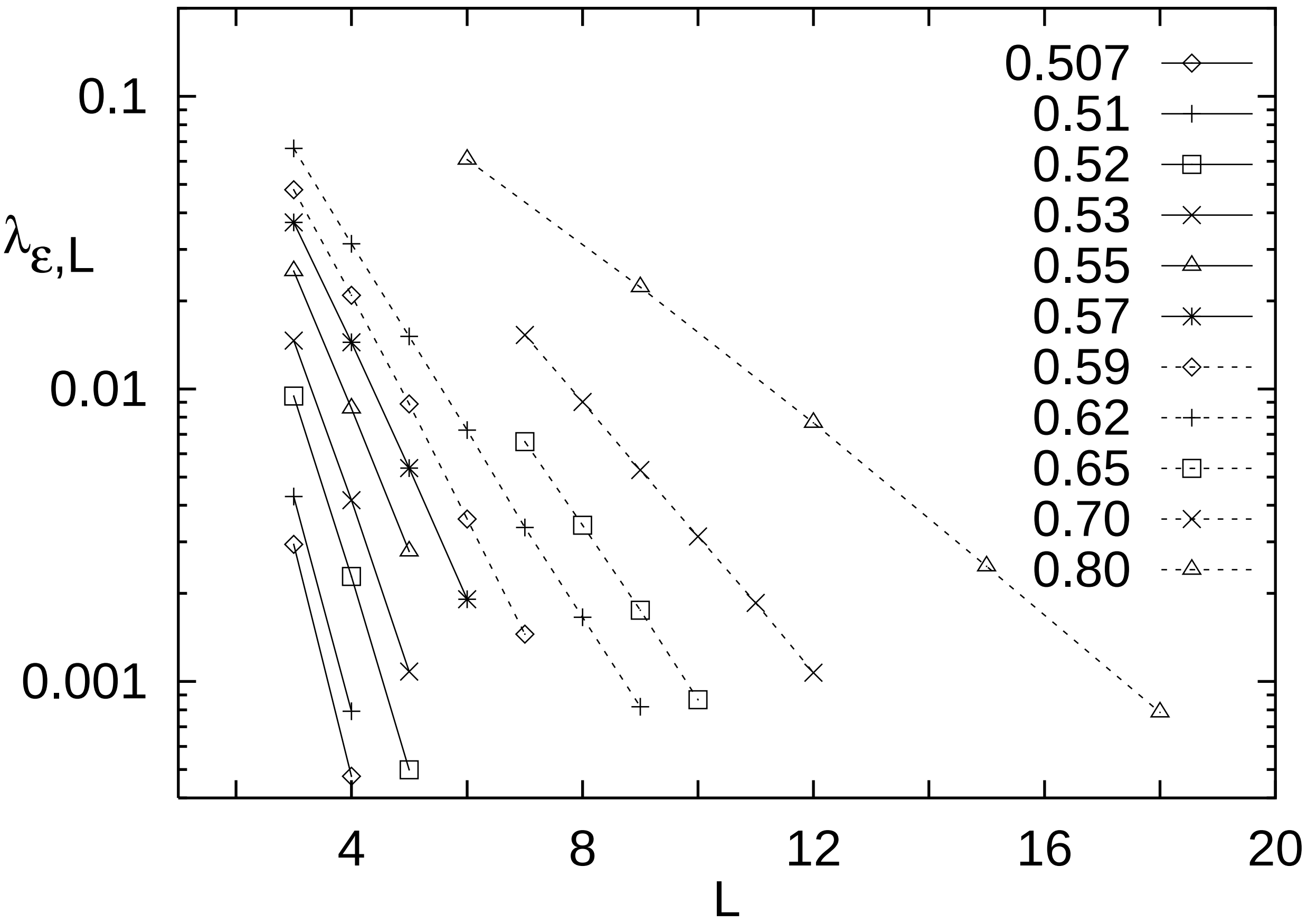

Fig.6

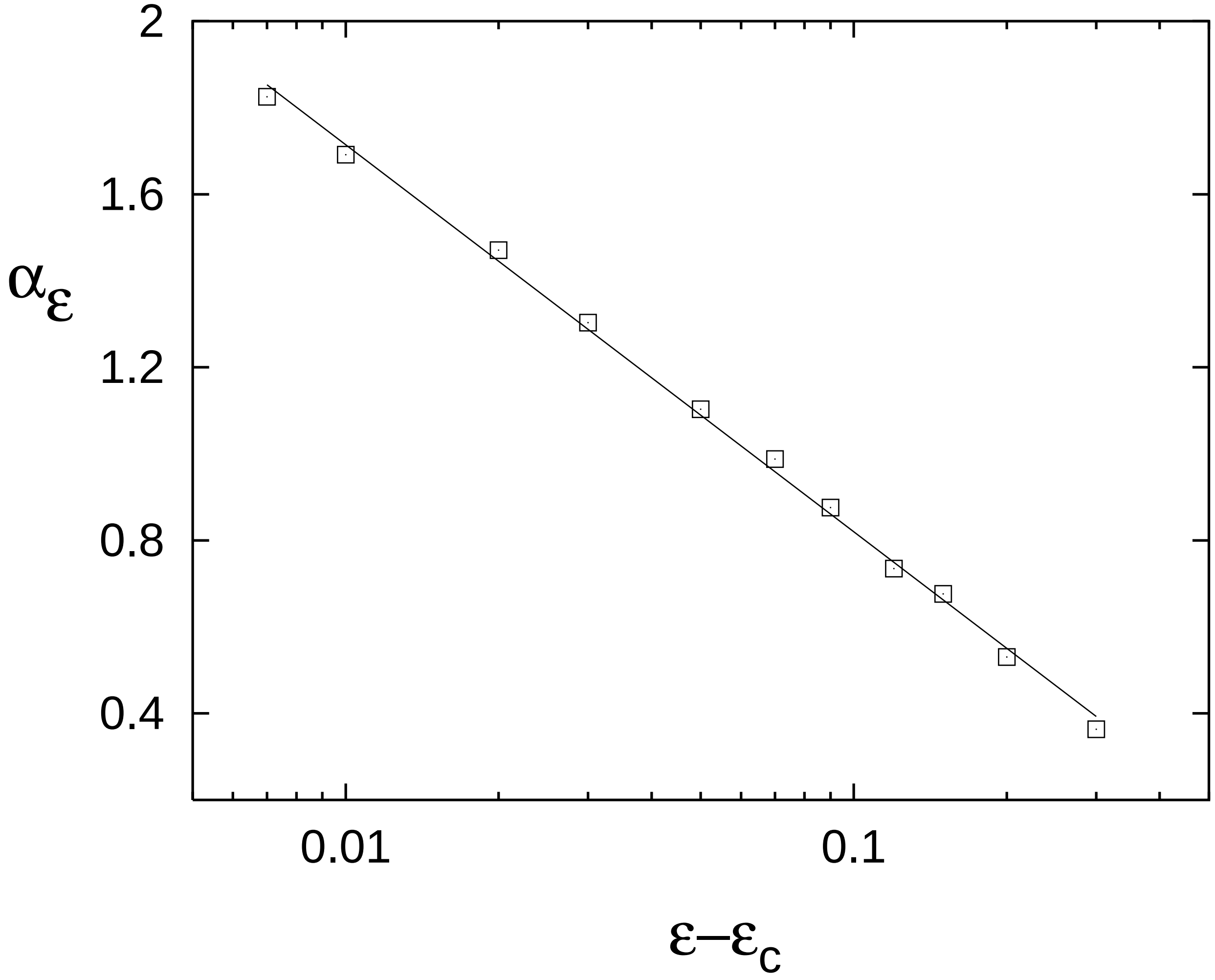

Fig.7

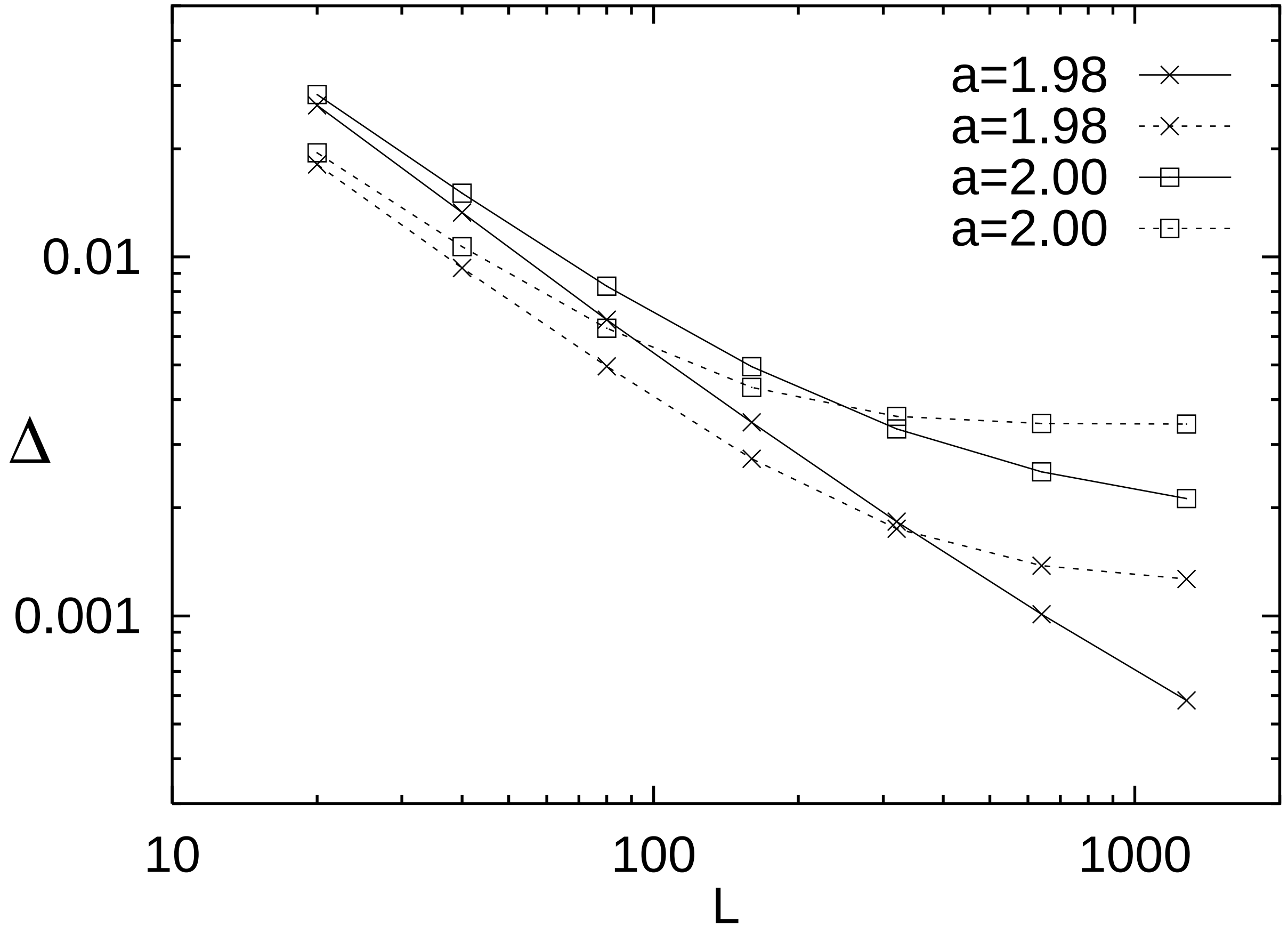

Fig.8a

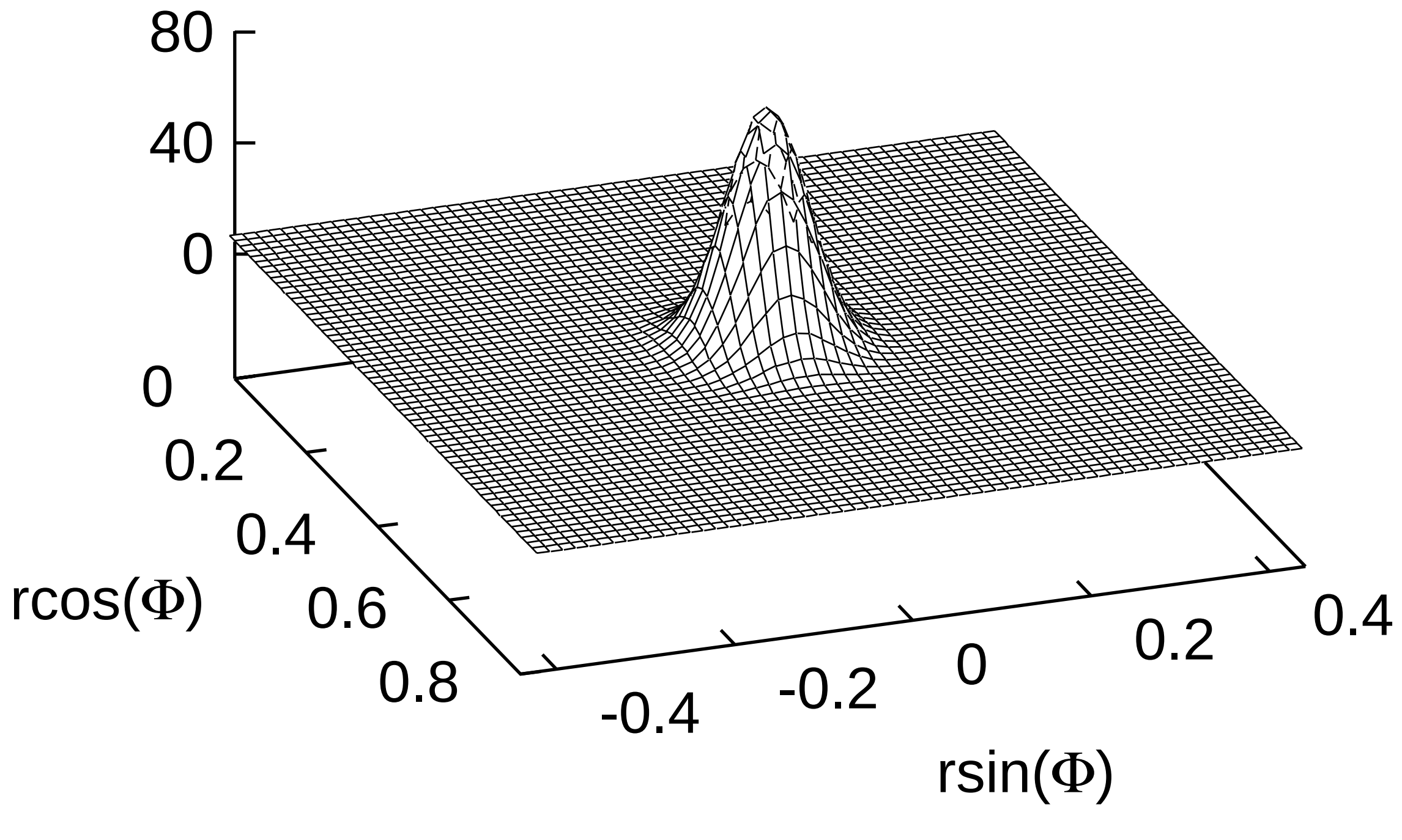

Fig.8b

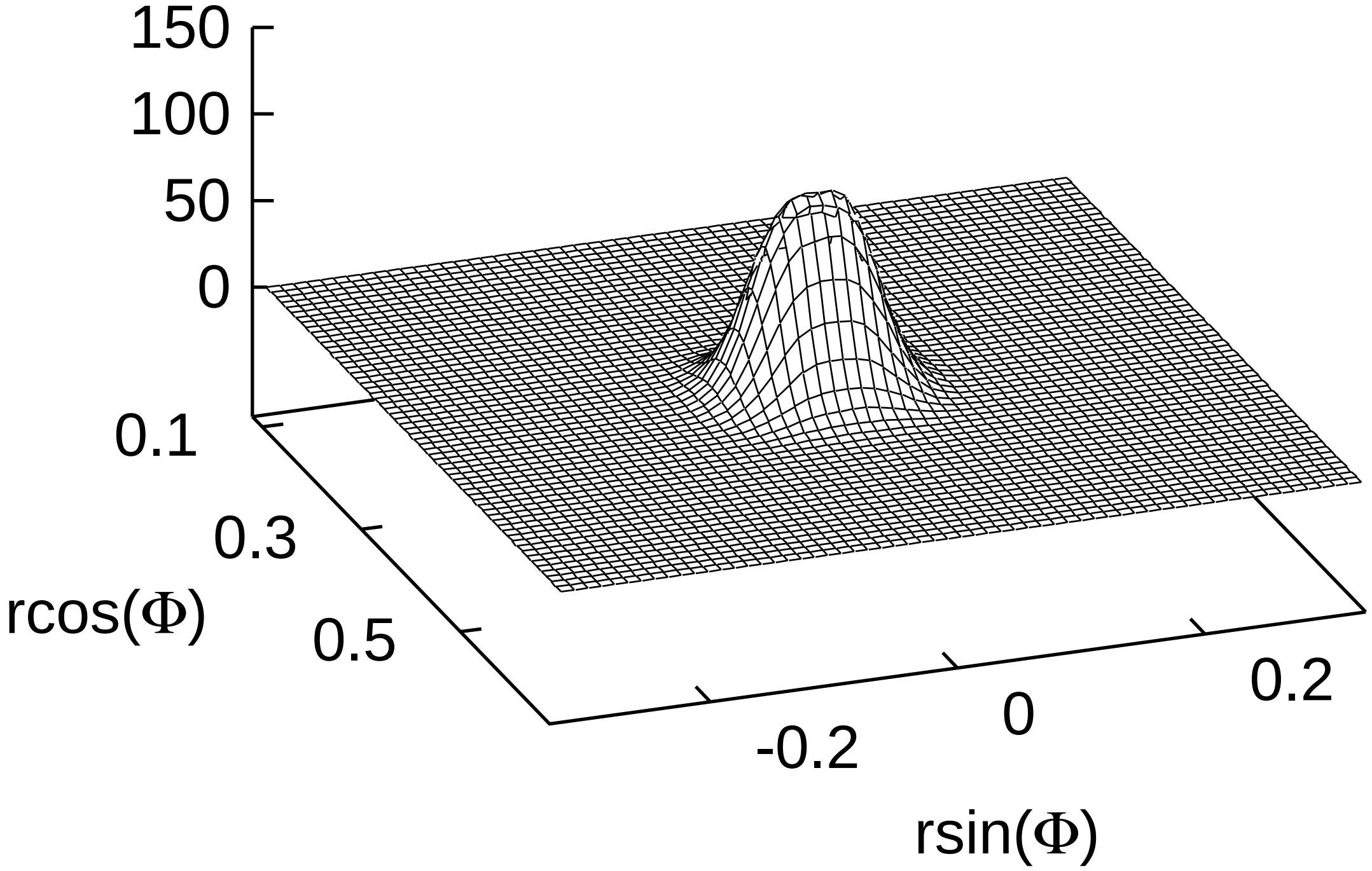

Fig.8c

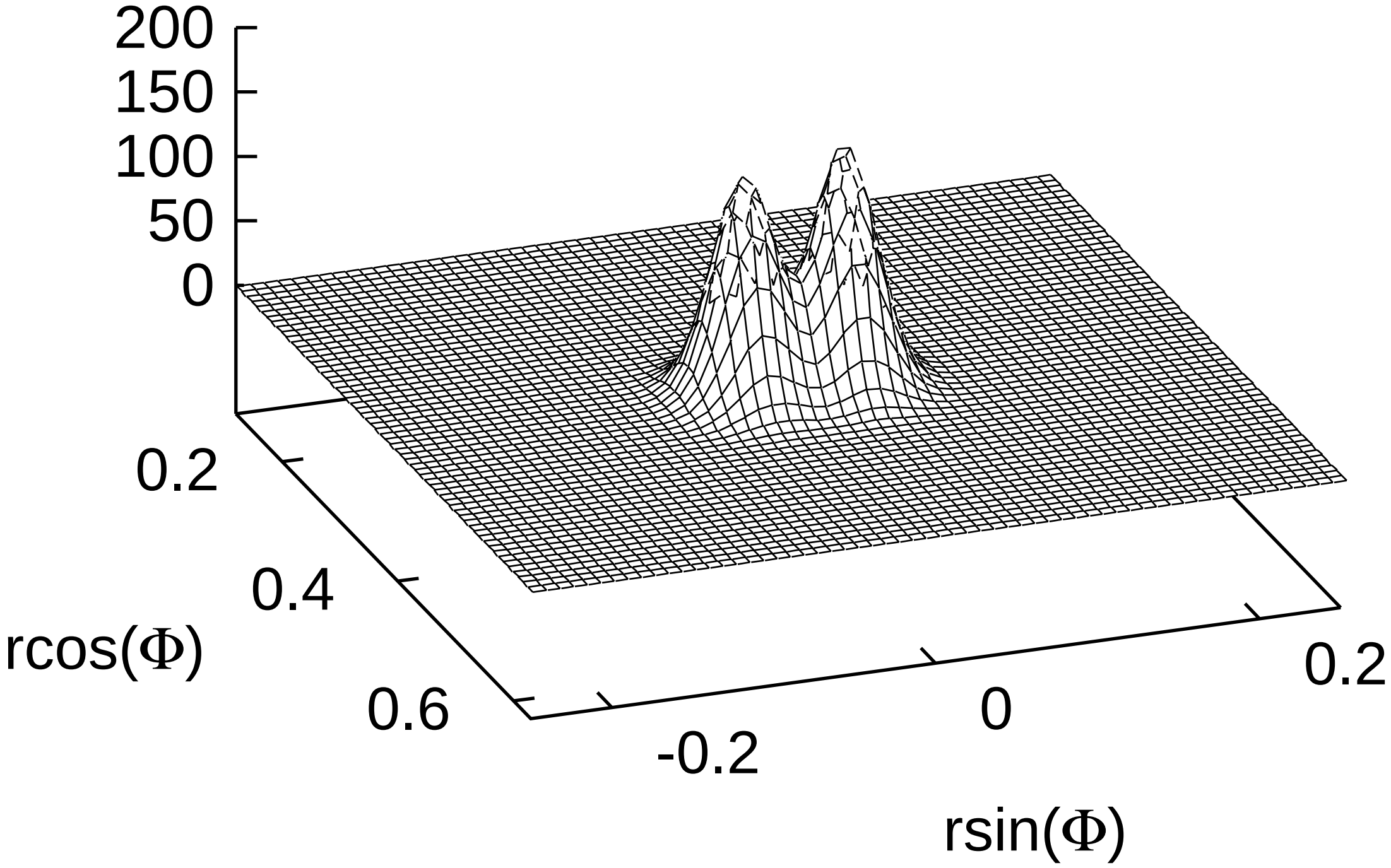

Fig.9a

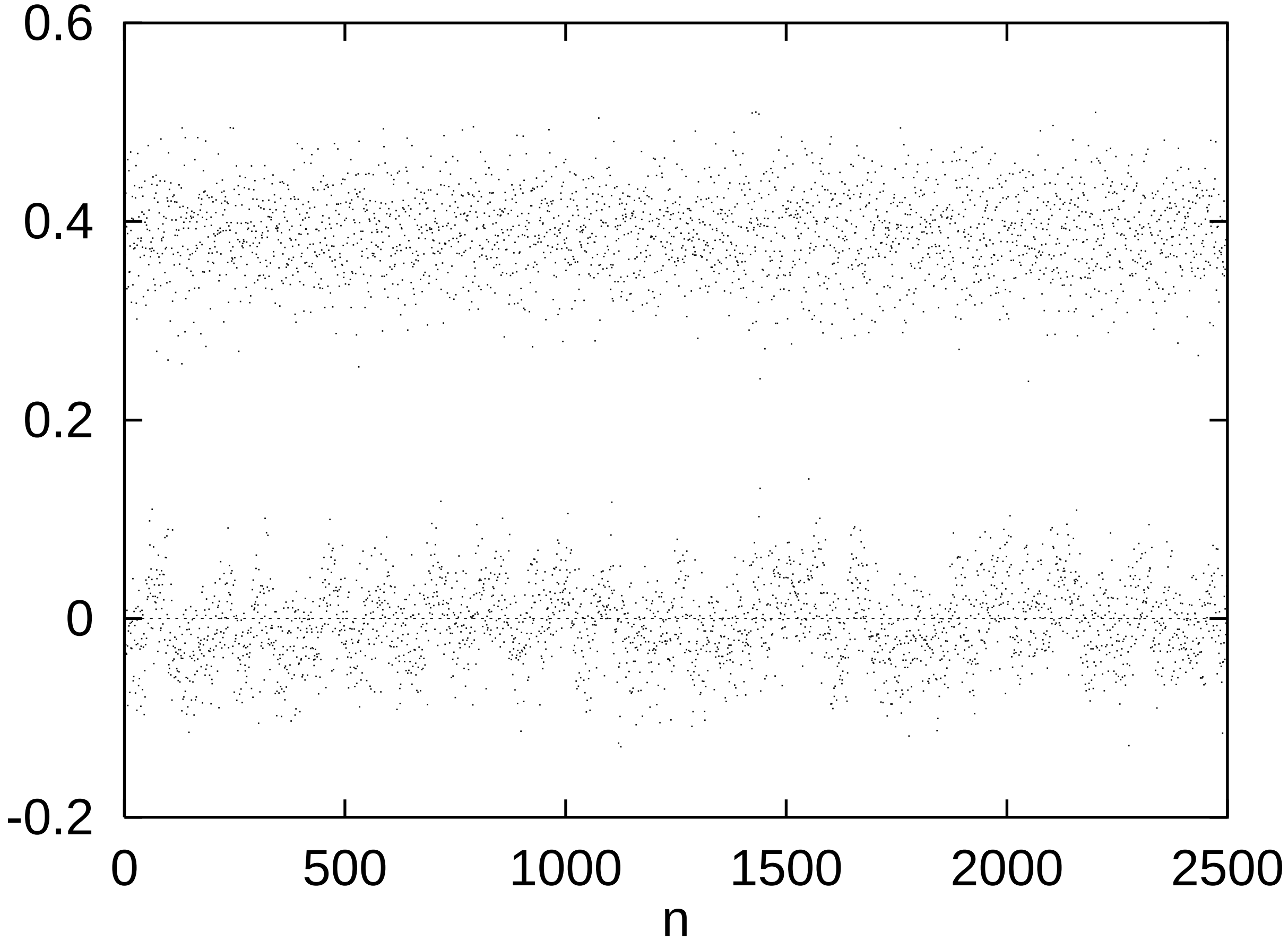

Fig.9b

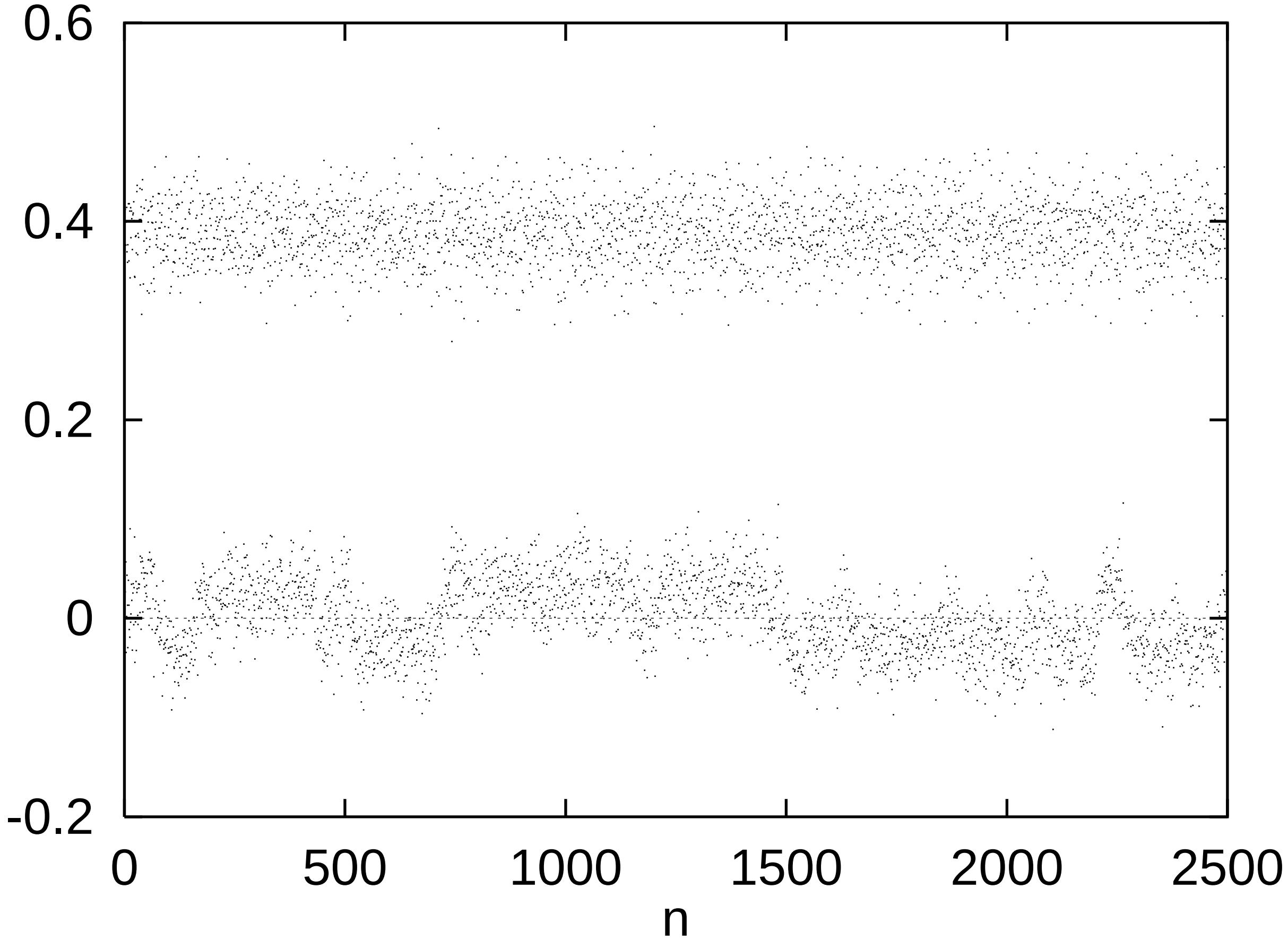

Fig.9c

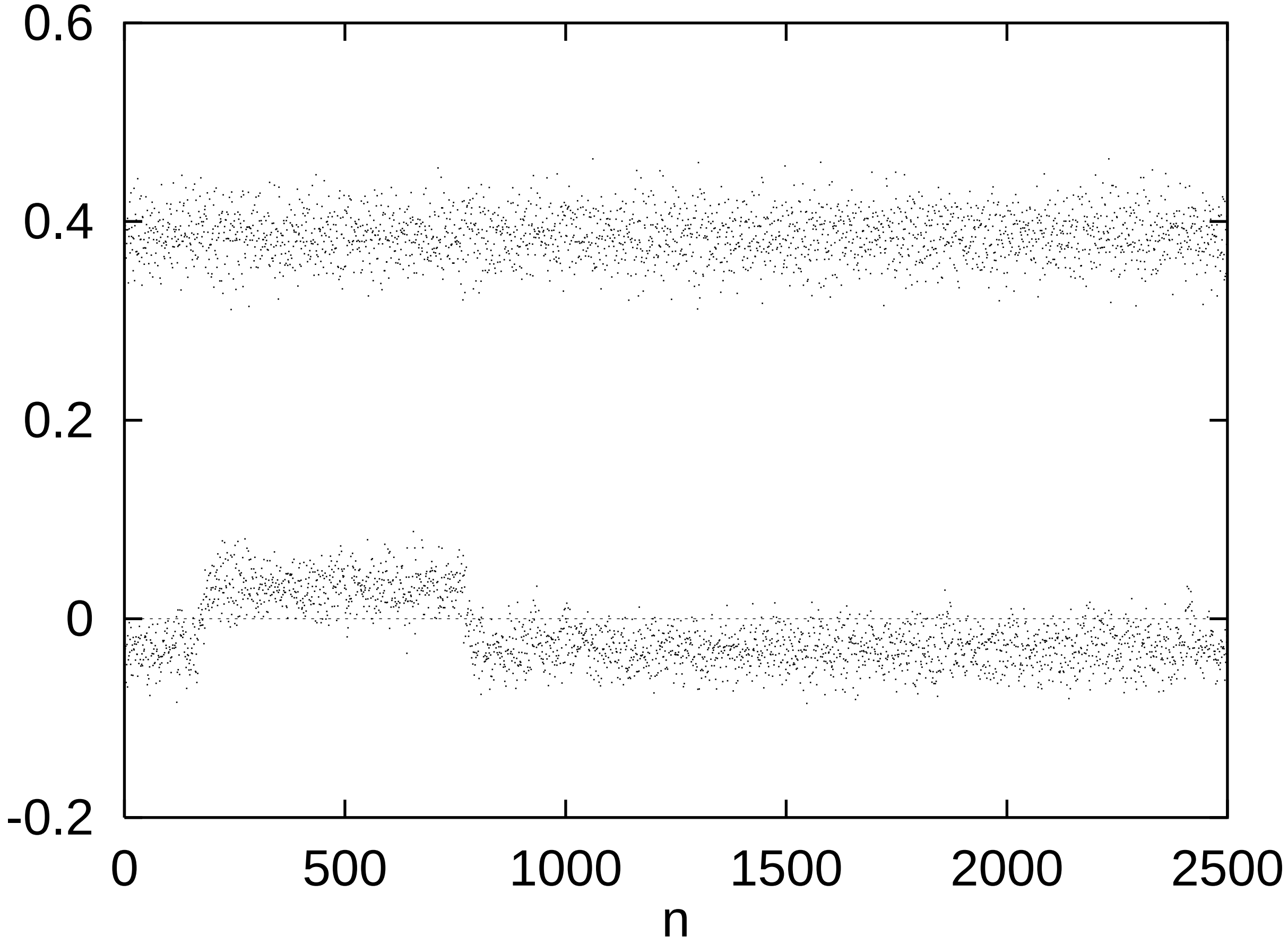

Fig.10a

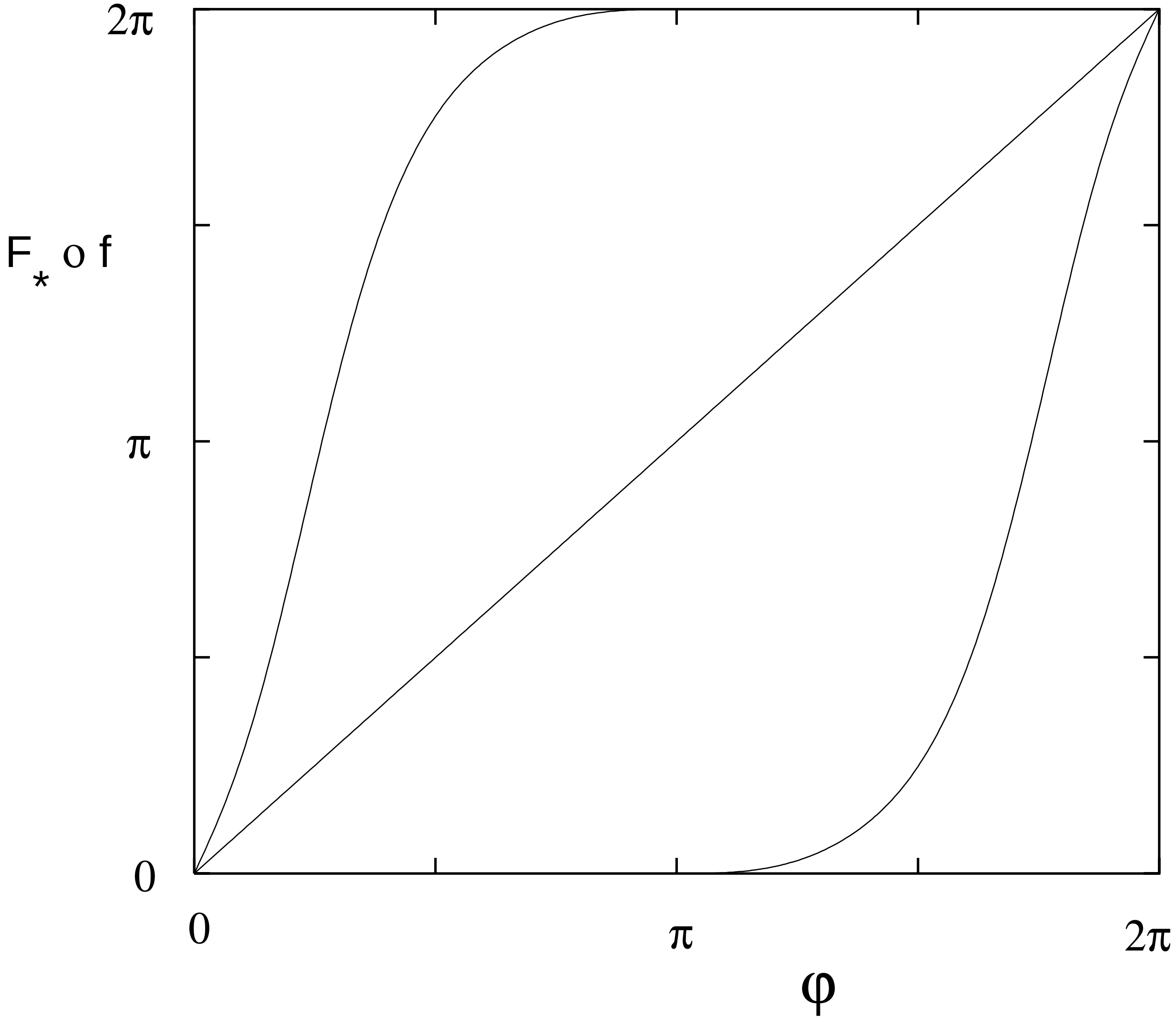

Fig.10b

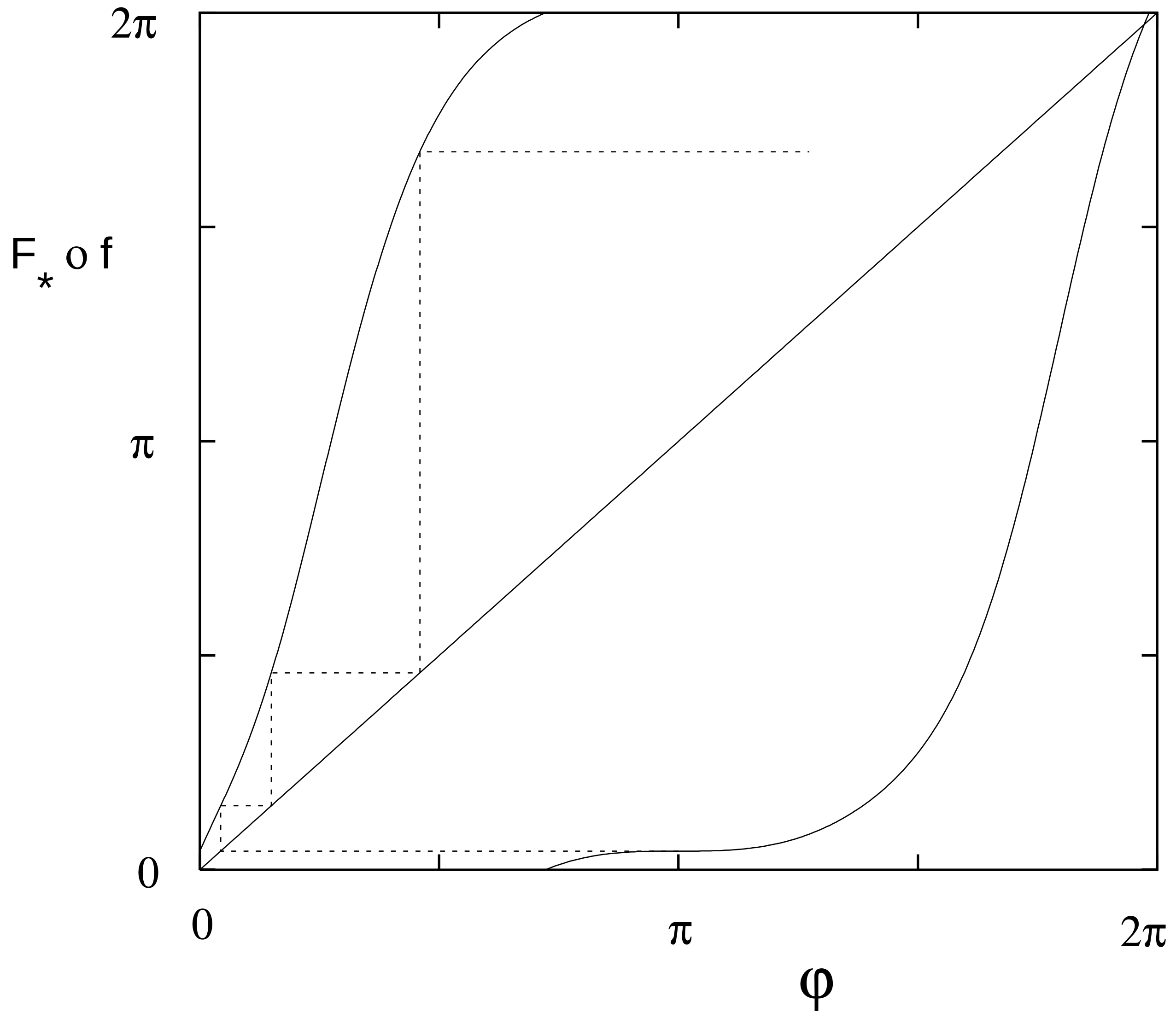

Fig.11a

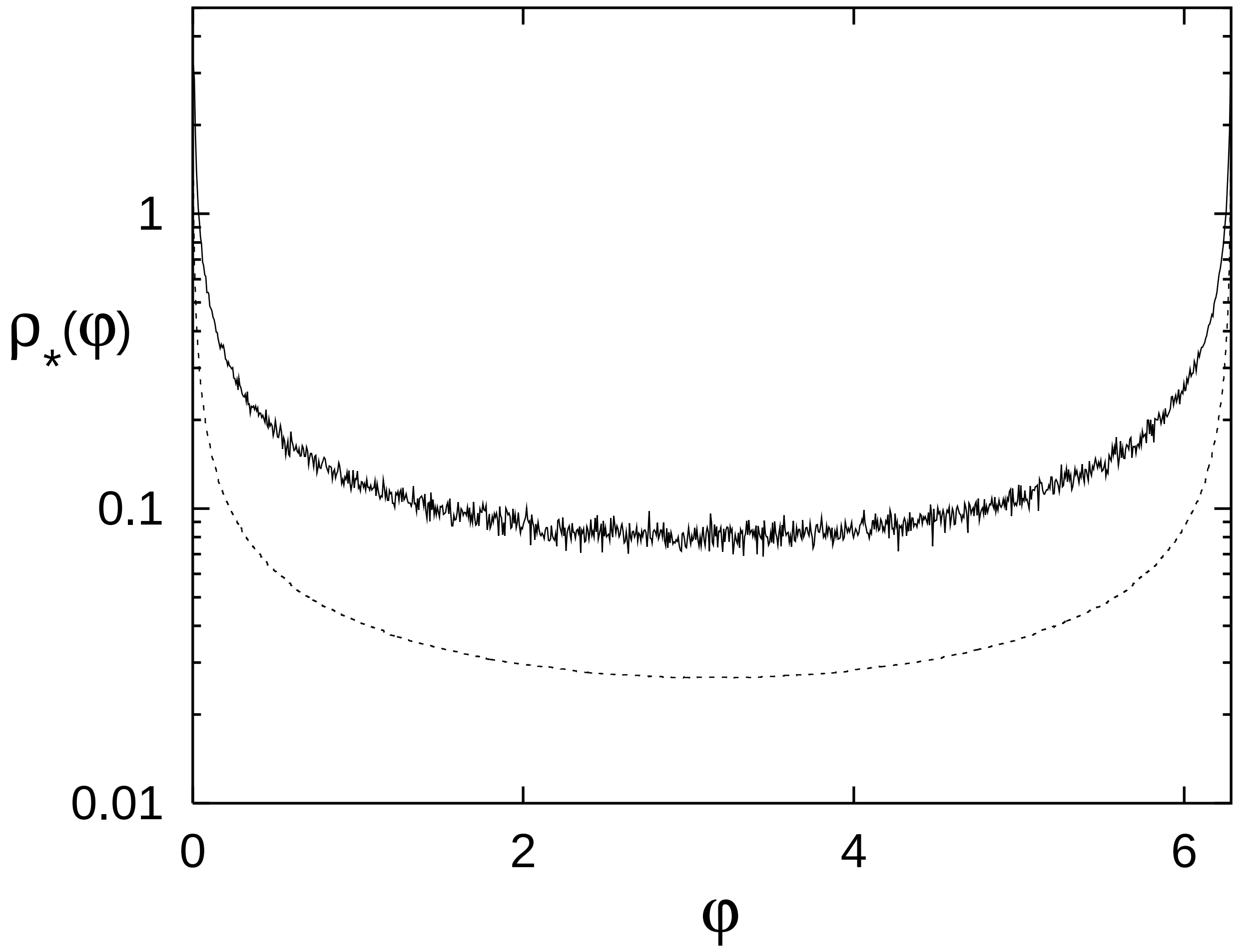

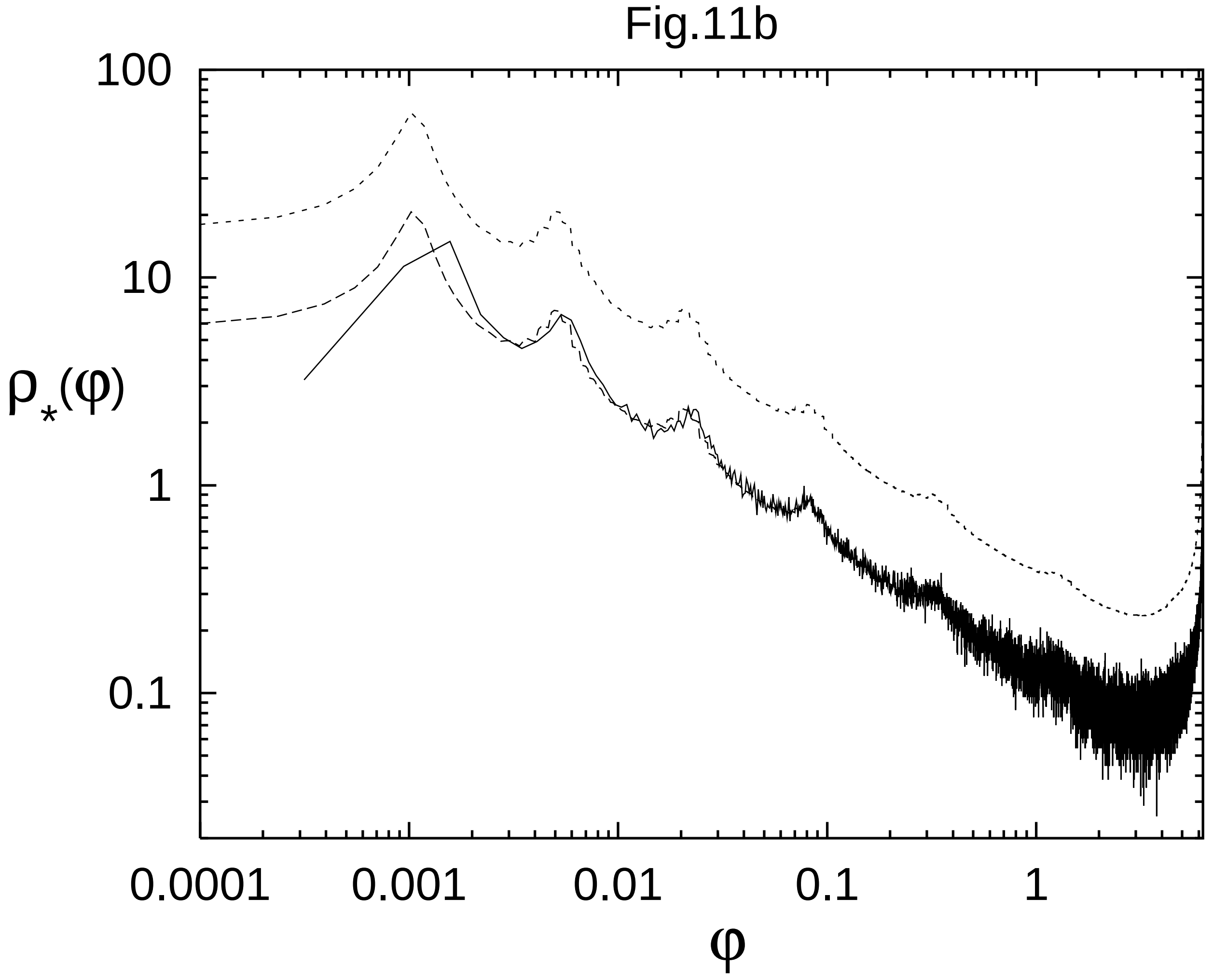

Fig.11c

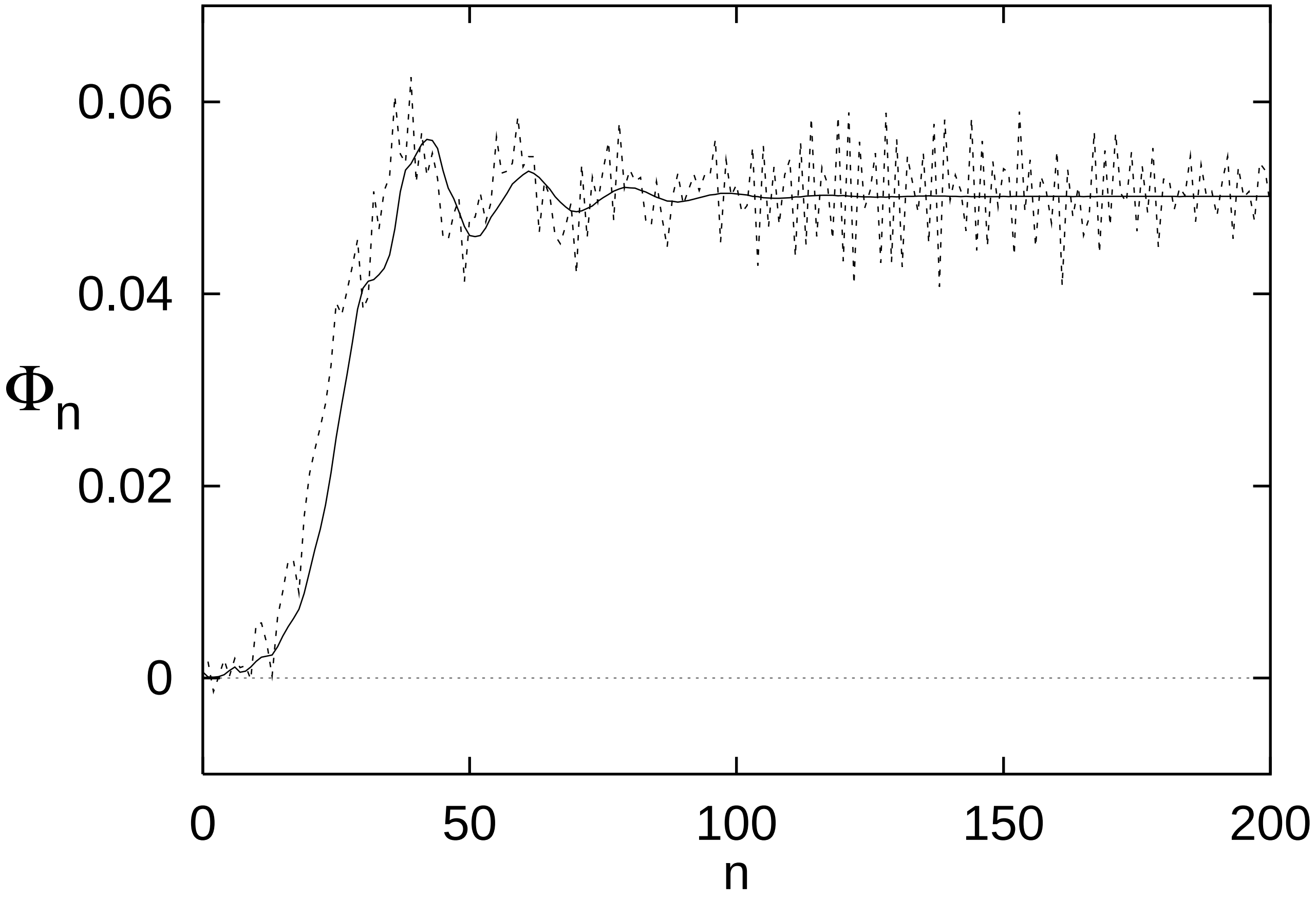

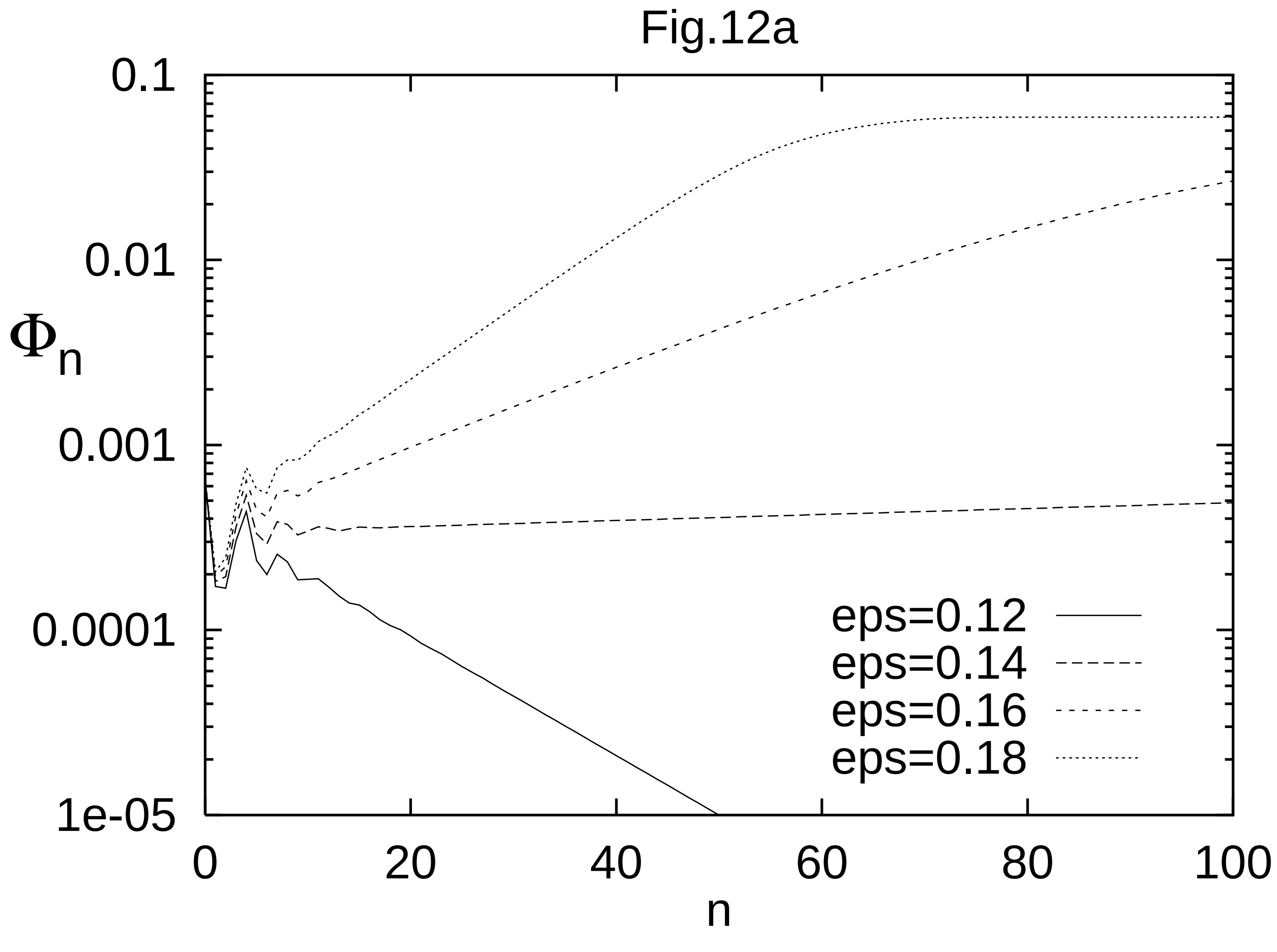

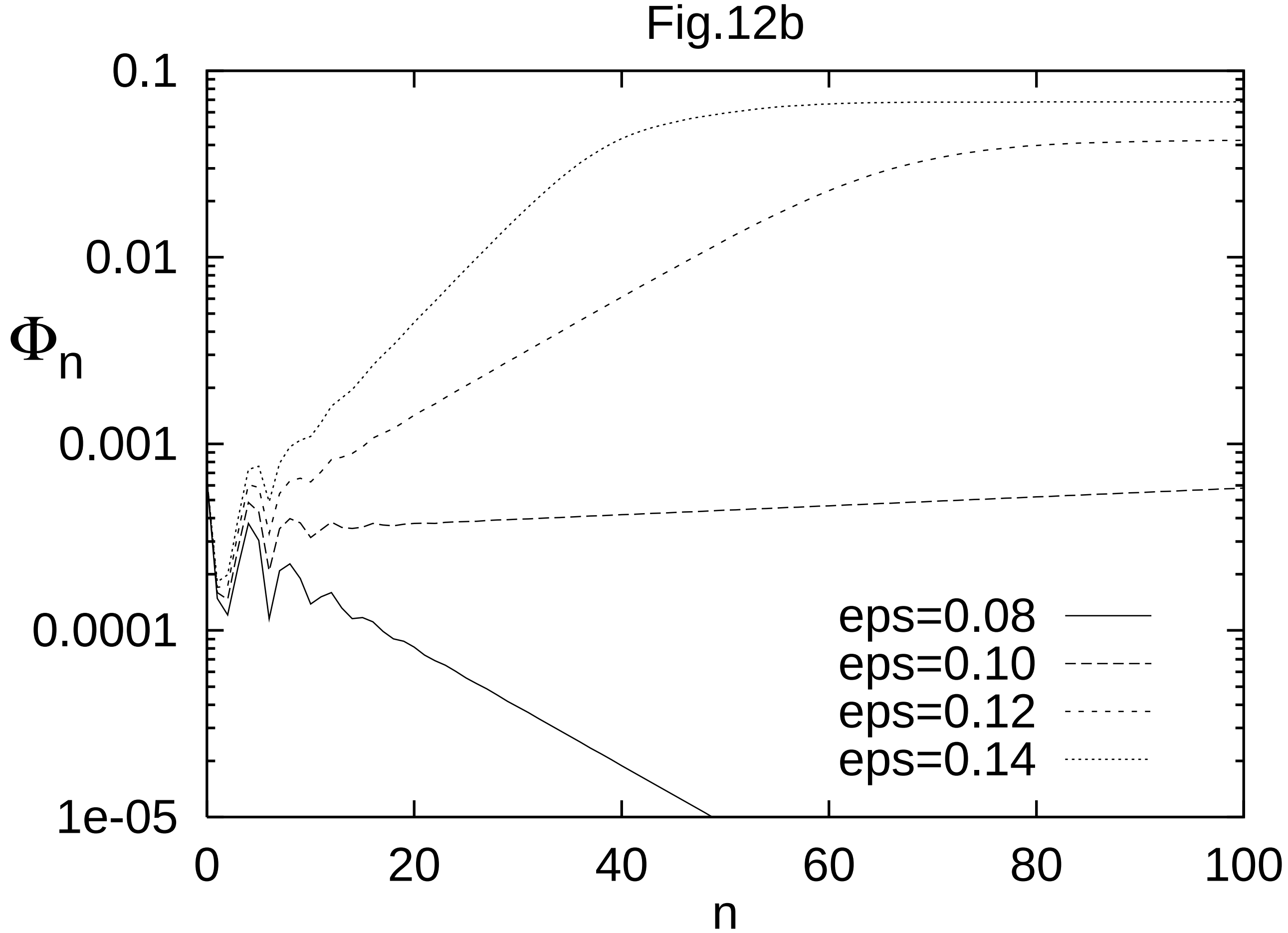

Fig.12c

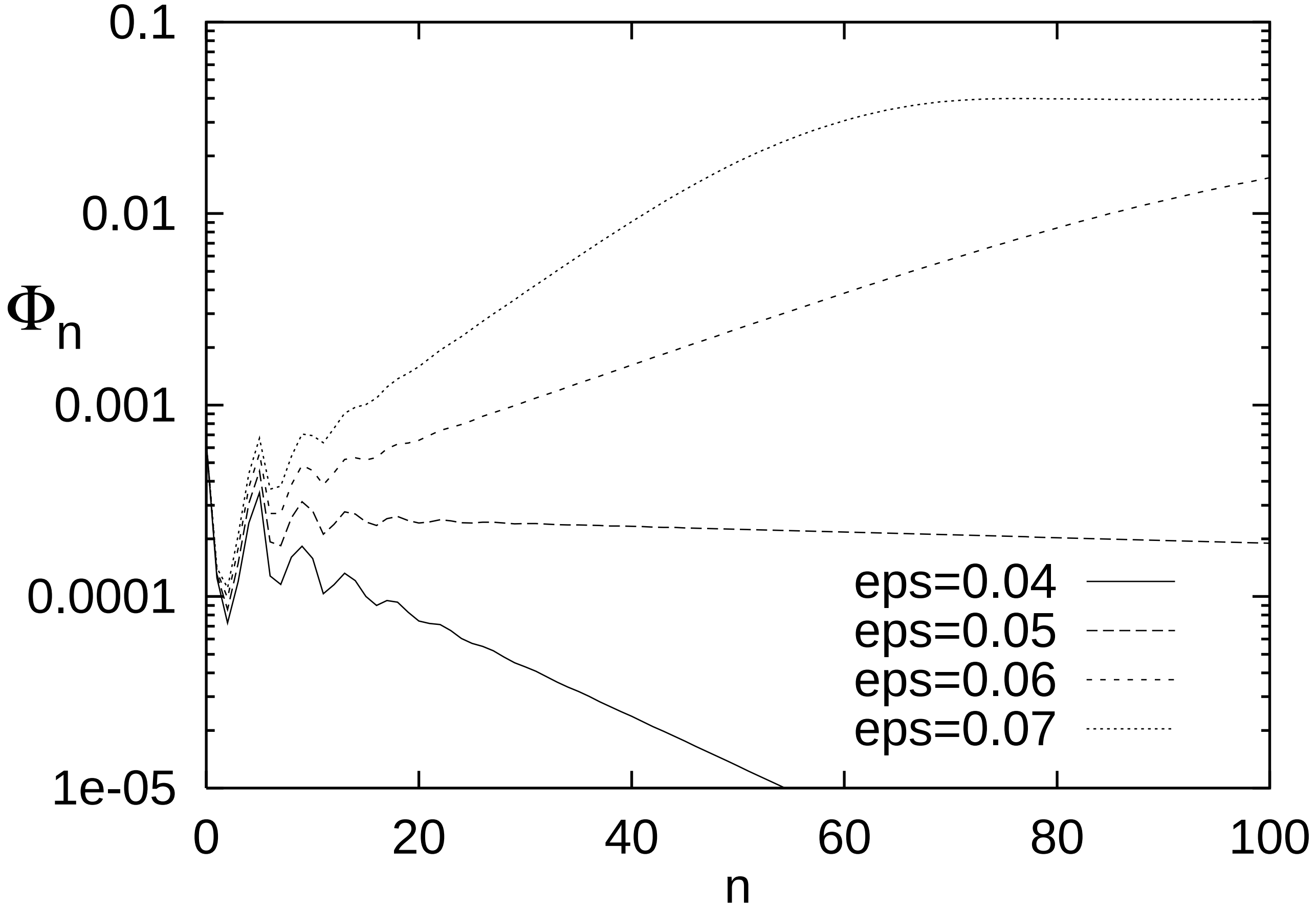

Fig.12d

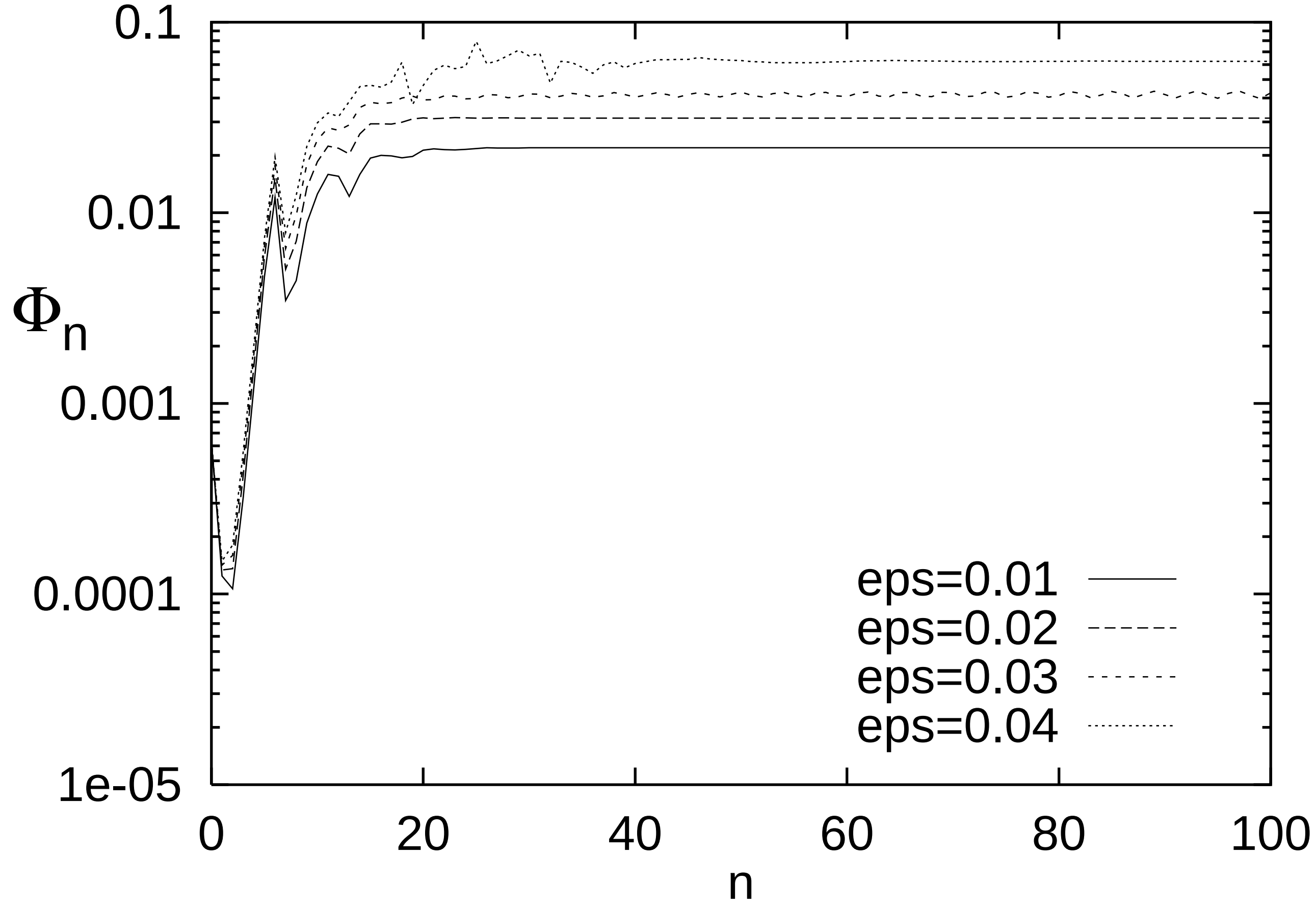

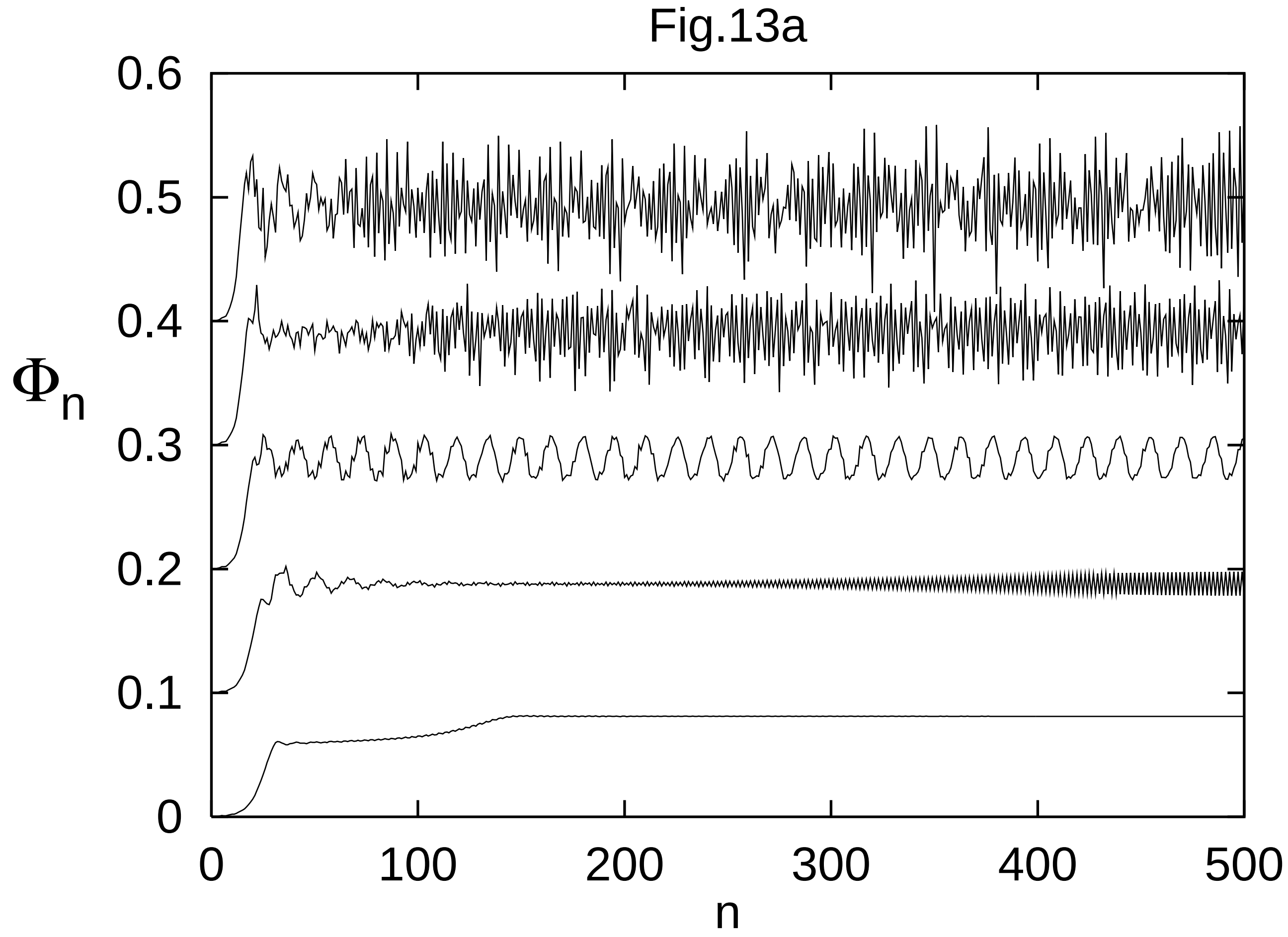

Fig.13b

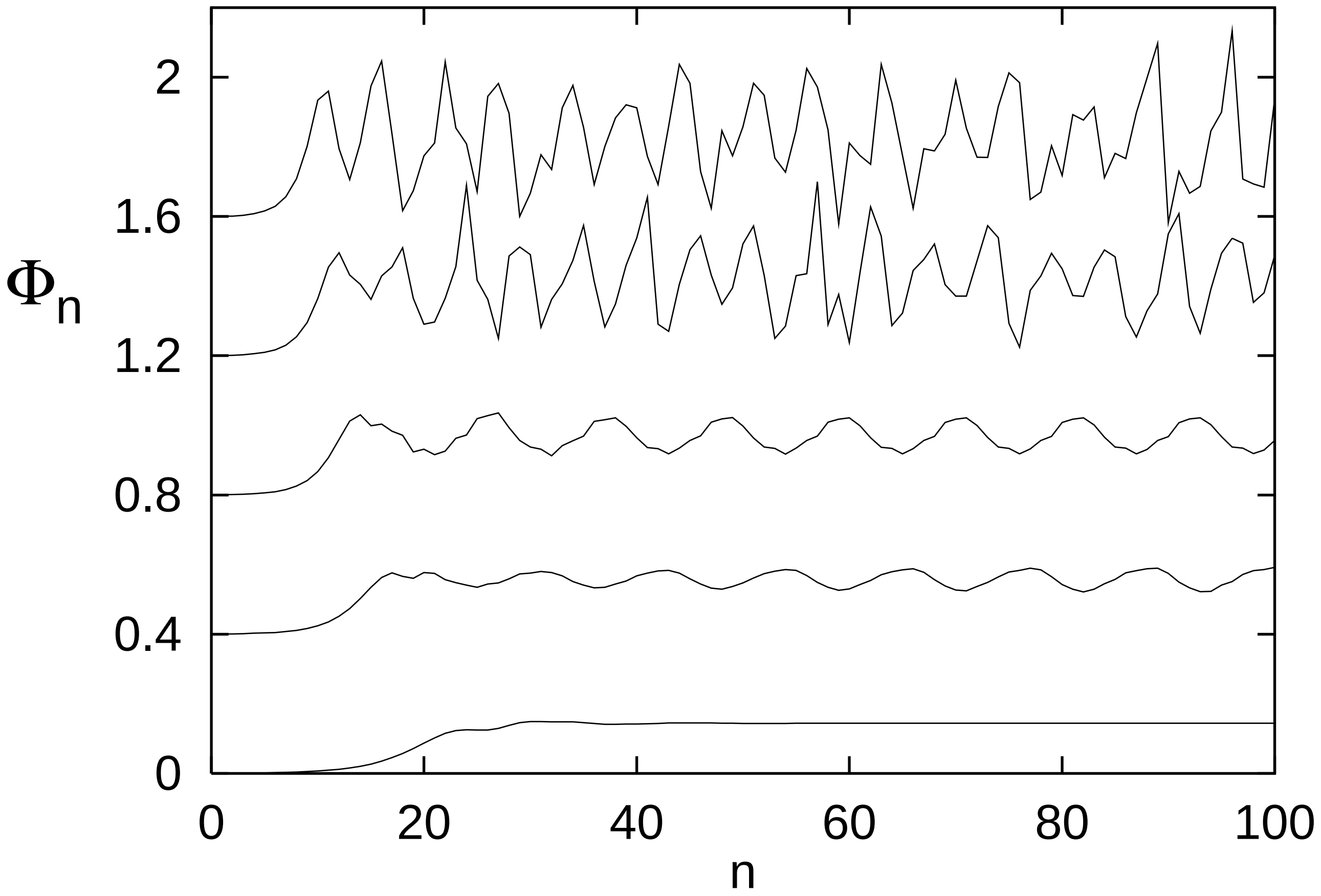